\documentclass[aps,prb,twocolumn,superscriptaddress]{revtex4-2}

\usepackage{amsmath}
\usepackage{amsfonts}
\usepackage{siunitx}
\usepackage{xcolor}
\usepackage{graphicx}
\usepackage{bm}
\usepackage[caption=false]{subfig}
\usepackage[export]{adjustbox}

\usepackage[utf8]{inputenc}
\usepackage[english]{babel}

\begin{document}

\title{Quantum magnetic oscillations in the absence of closed electron
trajectories}


\author{Z. E. Krix}
\email[]{z.krix@unsw.edu.au}
\affiliation{School of Physics, The University of New South Wales, Sydney, 2052, Australia}
\affiliation{Australian Research Council Centre of Excellence in
Low-Energy Electronics Technologies, The University of New South Wales, Sydney 2052, Australia}

\author{O. A. Tkachenko}
\affiliation{Rzhanov Institute of Semiconductor Physics, Novosibirsk, 630090, Russia}

\author{V. A. Tkachenko}
\affiliation{Rzhanov Institute of Semiconductor Physics, Novosibirsk, 630090, Russia}
\affiliation{Novosibirsk State University, Novosibirsk, 630090, Russia}

\author{D. Q. Wang}
\affiliation{School of Physics, The University of New South Wales, Sydney, 2052, Australia}
\affiliation{Australian Research Council Centre of Excellence in
Low-Energy Electronics Technologies, The University of New South Wales, Sydney 2052, Australia}

\author{O. Klochan}
\affiliation{School of Physics, The University of New South Wales, Sydney, 2052, Australia}
\affiliation{Australian Research Council Centre of Excellence in
Low-Energy Electronics Technologies, The University of New South Wales, Sydney 2052, Australia}

\author{A. R. Hamilton}
\affiliation{School of Physics, The University of New South Wales, Sydney, 2052, Australia}
\affiliation{Australian Research Council Centre of Excellence in
Low-Energy Electronics Technologies, The University of New South Wales, Sydney 2052, Australia}

\author{O. P. Sushkov}
\affiliation{School of Physics, The University of New South Wales, Sydney, 2052, Australia}
\affiliation{Australian Research Council Centre of Excellence in
Low-Energy Electronics Technologies, The University of New South Wales, Sydney 2052, Australia}

\date{\today}

\begin{abstract}
Quantum magnetic oscillations in crystals are typically understood in
terms of Bohr-Sommerfeld quantisation, the frequency of oscillation is
given by the area of a closed electron trajectory. However, since the
1970s, oscillations have been observed with frequencies that do not
correspond to closed electron trajectories and this effect has remained
not fully understood. Previous theory has focused on explaining the
effect using various kinetic mechanisms, however, frequencies without a
closed electron orbit have been observed in equilibrium and so a kinetic
mechanism cannot be the entire story. In this work we develop a theory
which explains these frequencies in equilibrium and can thus be used to
understand measurements of both Shubnikov-de Haas and de Haas-van Alphen
oscillations. We show, analytically, that these frequencies arise due to
multi-electron correlations. We then extend our theory to explain a
recent experiment on artificial crystals in GaAs two-dimensional
electron gases, which revealed for the first time magnetic oscillations
having frequencies that are half of those previously observed. We show
that the half-frequencies arise in equilibrium from single-particle
dynamics with account of impurities. Our analytic results are reinforced
by exact numerics, which we also use clarify prior works on the kinetic
regime.
\end{abstract}

\maketitle

\section{Introduction}

Quantum oscillations in a magnetic field have a long history;  both the
Shubnikov–de Haas and de Haas–Van Alphen effects belong to textbooks
\cite{abrikosov_fundamentals_2017}. In these effects the resistance or
the magnetisation are periodic functions of inverse magnetic field with
``frequency'' equal to the area of the Fermi surface. Quantum magnetic
oscillations become slightly more complex in the magnetic breakdown
regime, where tunneling between different Fermi surfaces is possible and
results in multiple additional frequencies. This regime was discovered
in quantum oscillations in magnesium \cite{priestley_experimental_1963}
and was studied in numerous experiments after. Quantum magnetic
oscillations are an important experimental tool for the study of solids,
in part because of their relation to the size of the Fermi surface. They
are used in topological materials \cite{alexandradinata_geometric_2017,
alexandradinata_fermiology_2023}, in cuprates \cite{kunisada2020,
kurokawa_unveiling_2023} in twisted bilayer graphene
\cite{de_vries_kagome_2023} and in many other systems. In spite of the
very long history of studies, one type of quantum magnetic oscillations,
the type that we call ``non-Onsager'' throughout this work, has not been
fully understood theoretically. Moreover, recent experiments on
two-dimensional artificial crystals built on a GaAs two-dimensional
electron gas (2DEG) have discovered a new type of magnetic oscillation
effect: half-frequency non-Onsager
oscillations~\cite{wang_formation_2023}. The goal of the present work is
to explain the non-Onsager effects and to develop a theory which
accounts for both their frequency and the temperature dependence of
their amplitude.

\begin{figure}[ht]
    \centering
    \includegraphics[width=0.4\textwidth]{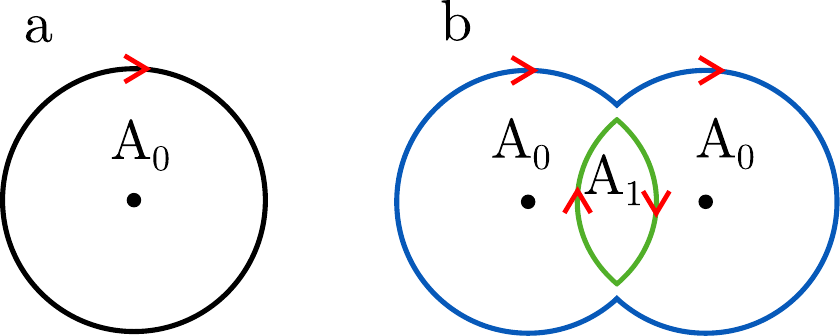}
    \caption{Panel a: simple Fermi surface and the corresponding circular trajectory in magnetic field.
    The corresponding frequency of quantum oscillations is $A_0$, the area of the Fermi surface.
    Panel b: additional closed semiclassical trajectories that arise due to quantum jumps
    induced by Bragg scattering from the periodic potential. The corresponding frequencies of quantum 
    oscillations are $A_1$, $2A_0-A_1$, $A_0+A_1$. 
   }
    \label{fig:intro}
\end{figure}

Standard quantum magnetic oscillations are described well by Onsager's
semi-classical picture \cite{onsager_interpretation_1952}: an electron
in a magnetic field undergoes periodic motion, orbiting along the Fermi
surface (see Fig. \ref{fig:intro}a). Quantum oscillations of resistivity
and magnetisation arise due to Bohr-Sommerfeld quantization of this
motion. The oscillations are periodic in inverse magnetic field and the
oscillation frequency is proportional to the cross-sectional area of the
Fermi surface, which we denote by $A_0$ for a circular Fermi surface.

When an electron gas is submerged in a periodic potential, magnetic
breakdown can arise in addition to simple Bohr-Sommerfeld quantization.
The potential can have a chemical origin, such as the lattice of a
crystal, or it can have an artificial origin, such as an external
modulation in a 2DEG~\cite{wang_two-dimensional_2020}. Magnetic
breakdown is a quantum correction to the simple Onsager picture, and was
discovered in magnesium in the 1960s \cite{priestley_experimental_1963}.
The origin of the effect is illustrated in panel b of
Fig.\ref{fig:intro}. Two electron orbits (blue circles) arise due to
diffraction on the periodic potential. When the orbits intersect there
is a probability that an electron will quantum mechanically jump from
one circle to another. Hence, in addition to $A_0$, quantum oscillations
manifest $A_1$, $2A_0-A_1$, and $A_0+A_1$ frequencies, where $A_{1}$ is
the ``elliptical" orbit drawn in green. In essence this is the same
Onsager picture, only the number of possible trajectories is increased
due to quantum jumps. Therefore, we call all these oscillations
(frequencies) Onsager quantum oscillations. This is the magnetic
breakdown regime. It is worth noting that the notion of Fermi surface is
somewhat ambiguous in this regime. Multiple Onsager quantum oscillations
have been observed in numerous experiments.

The multiple Onsager quantum oscillations are not puzzling, what is
puzzling is that magnetic oscillations with frequency $A_{0}-A_{1}$ have
also been observed! Of course, the moon-shaped ``trajectory'' with area
$A_{0}-A_{1}$ exists in Fig. \ref{fig:intro}b. However, an electron
cannot traverse along this ``trajectory'' since in a magnetic field the
electron must travel in the same direction along each circular arc. The
$A_{0}-A_{1}$ oscillations are beyond the Onsager picture; we refer to
them as non-Onsager oscillations. Non-Onsager oscillations can be more
complex than just $A_{0} - A_{1}$,  however, they share the same
property of not being traversable by an electron. Below, we enumerate
some observations relating to these kinds of oscillations.

{\bf (i)} Non-Onsager oscillations $A_{0} - A_{1}$ have been observed in
transport experiments, first in Magnesium \cite{stark_interfering_1974,
morrison_two-lifetime_1981}, then in laterally modulated, GaAs 2DEGs
\cite{gerhardts_novel_1989, deutschmann_quantum_2001}, and later in
organic metals \cite{meyer_high-field_1995, audouard_quantum_2012}. 
A common feature is that they decay with temperature much more slowly
than Onsager oscillations. In semiconductors these oscillations are 
called Weiss oscillations or ``commensurate oscillations''.

{\bf (ii)} In the limit of very large wavelength of the external
potential the ``commensurate oscillations'' in transport have been
explained on the basis of the single particle kinetic equation as an
essentially non-equilibrium effect \cite{morrison_two-lifetime_1981,
beenakker_guiding-center-drift_1989, winkler_landau_1989,
mirlin_weiss_1998}. The theoretical work Ref.\cite{kaganov_coherent_1983}
explicitly claims that non-Onsager oscillations can appear only in
resistivity and not in magnetisation. In a sense, the single particle
non-equilibrium kinetic equation mechanism implies that ``commensurate
oscillations'' are magnetic oscillations, but they are not quantum
magnetic oscillations.

{\bf (iii)} Contrary to the single electron kinetic mechanism,
non-Onsager magnetic oscillations have been also observed in
magnetisation measurements in magnesium \cite{eddy_haas---van_1982} and
in organic metals \cite{audouard_quantum_2012}. This means the effect
exists in equilibrium as well. It does not mean that the non-equilibrium
kinetic equation mechanism \cite{morrison_two-lifetime_1981,
beenakker_guiding-center-drift_1989,
winkler_landau_1989,mirlin_weiss_1998} is wrong, however, it does mean
that kinetics is only a part of the story.

{\bf (iv)} There is another twist related to 2D organic metals. It is
well known that with variations of the magnetic field in 2D gated
systems the electron density is constant while the chemical potential is
oscillating. There is a claim in literature~\cite{fortin_frequency_1998}
that non-Onsager quantum oscillations arise as combinatorial frequencies
between the chemical potential oscillations and oscillations in the
grand potential, so non-Onsager frequencies cannot exist at constant
chemical potential. However, it has been pointed out
later~\cite{gvozdikov_haas--van_2002} that the ``oscillating chemical
potential'' explanation cannot be correct since non-Onsager frequencies
have also been observed in magnesium at constant chemical potential.
Moreover, the observed non-Onsager quantum oscillations decay with
temperature very slow, slower than the main oscillations $A_0$. At the
same time the mechanism related to oscillations of the chemical
potential results in extremely fast decay with temperature.

To summarise points {\bf (i)-(iv)}, there is no consistent understanding
of the mechanism of the non-Onsager magnetic oscillations. In our recent
transport experiment \cite{wang_formation_2023} on a triangular lattice
of anti-dots in a GaAs 2DEG we observed the non-Onsager oscillations
$A_0-A_1$ as well as other non-Onsager and Onsager oscillations that
have not been previously observed. Moreover, we have also discovered
half-frequency non-Onsager and Onsager oscillations with frequencies
such as $(A_{0} - A_{1})/2$ and $(A_{0} + A_{1})/2$. Half-frequency
oscillations have not been observed in any previous experiments and have
not been discussed in any previous theoretical works. The important
observation that sheds light on the mechanism is that amplitudes of
$A_{0} - A_{1}$ and $(A_{0} - A_{1})/2$ oscillations decay with
temperature much slower than the amplitudes of Onsager oscillations.

The purpose of this paper is to explain these non-Onsager quantum
oscillations. First, we resolve the controversy of ``known'' non-Onsager
oscillations (``known'' meaning prior to Ref.
\cite{wang_formation_2023}). We demonstrate that there are two
mechanisms: (i) single particle kinetics, and (ii) multi-electron
correlations due to the Coulomb electron-electron interaction. Both
mechanisms contribute to the Shubnikov–de Haas effect (transport) and
only the second mechanism contributes to the de Haas–Van Alphen effect
(equilibrium). We calculate the temperature dependence of quantum
non-Onsager oscillations and hence explain why they survive up to
relatively high temperatures in spite of being fully quantum. Second, we
explain the recently discovered
half-frequencies~\cite{wang_formation_2023}. Here there are also two
mechanisms, both single particle: (i) single particle kinetics, (ii)
single particle dynamics with account of impurities. Both mechanisms
contribute to the Shubnikov–de Haas effect (transport) and only the
second mechanism contributes to the De Haas–Van Alphen effect
(equilibrium). The mechanism (ii) conceptually has a distant similarity
to the Al'tshuler-Aronov-Spivak effect
\cite{altshuler_aaronov-bohm_1981}.

The structure of the paper is the following. In section II we find
analytically energies and wave functions of an electron in superimposed
1D periodic potential and magnetic field. The main conclusion is that
the density of states (DOS) does not contain non-Onsager frequencies. In
Section III we solve numerically the problem of electron transport
through a region with modulated potential. We consider a 1D periodic
potential and a 2D triangular lattice. The numerical approach confirms
that non-Onsager frequencies are absent in DOS, but there are integer
and half-integer non-Onsager frequencies in the resistance. While
qualitatively the results for integer non-Onsager frequencies are
consistent with the previous picture of commensurate oscillations,
quantitatively our conclusions are somewhat different. The results of
this Section explain all known experimental data on the Shubnikov–de
Haas effect, but do not explain oscillations observed in the de Haas–Van
Alphen effect. In Section IV we calculate the thermodynamic potential of
an ideal Fermi gas in a superimposed 1D periodic potential and magnetic
field, this is a technical section. Section V addresses
electron-electron interactions and explains non-Onsager oscillations
observed in the de Haas–Van Alphen effect. The analysis includes the
temperature dependence of the amplitude of the oscillations. Section VI
is aimed at chemical potential oscillations in a gated 2DEG. This
mechanism leads to non-Onsager oscillations in free energy, however, it
does not explain the observed temperature dependence of the effect. In
Section VII we consider half-frequency non-Onsager oscillations due to
impurities. This effect conceptually has a distant similarity to the
Al'tshuler-Aronov-Spivak effect \cite{altshuler_aaronov-bohm_1981}.
Section VIII presents our conclusions.

\section{Motion in a one-dimensional periodic potential}

\subsection{Energy levels }

Our results are generic, however, for the sake of presentation we work
with a specific potential: the one-dimensional sinusoidal potential.

\begin{eqnarray}\label{u1D}
    U(x) = 2 W \cos(g x)
\end{eqnarray}

\begin{figure}[h]
    \centering
    \includegraphics[width=0.4\textwidth]{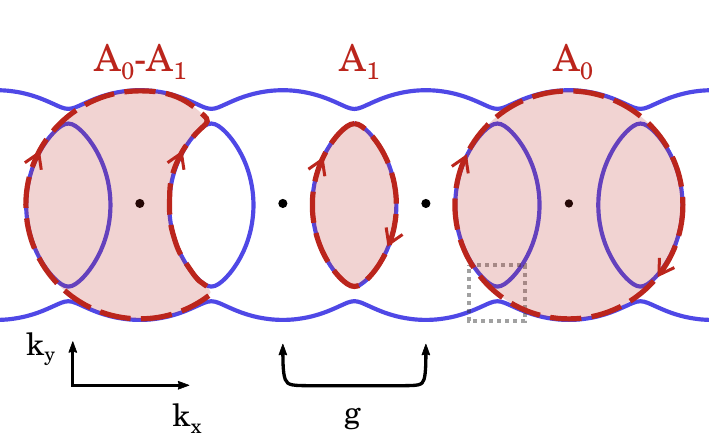}
    \caption{Blue lines show Fermi surface of a 2DEG with a one-dimensional periodic potential (Eqn. \ref{u1D}). Red lines show Semi-classical electron "trajectories" with arrows indicating the direction of motion.}
    \label{fig:gratingFermiSurfaceA0A1}
\end{figure}

This potential is imposed on a two-dimensional electron gas (2DEG) with
quadratic dispersion, $\varepsilon(\bm{p}) = \frac{p^2}{2m}$. We assume
that both the potential modulation and the magnetic field are small
compared to the Fermi energy, $W, \hbar\omega \ll \varepsilon_F$, so
that the semi-classical approximation is valid. The semi-classical
trajectory of a free electron is a circle in momentum space (Fig.
\ref{fig:intro}a); the sign of the magnetic field corresponds to
clockwise propagation. Diffraction from the potential, Eq. \ref{u1D},
makes the ``trajectory'' periodic in $k_{x}$ (see Fig.
\ref{fig:gratingFermiSurfaceA0A1}, the propagation remains clockwise).
Note that Fig. \ref{fig:gratingFermiSurfaceA0A1} corresponds to the case
$g / 2 < p_{F} < g$.

\begin{figure}[h]
    \centering
    \includegraphics[height=0.11\textwidth]{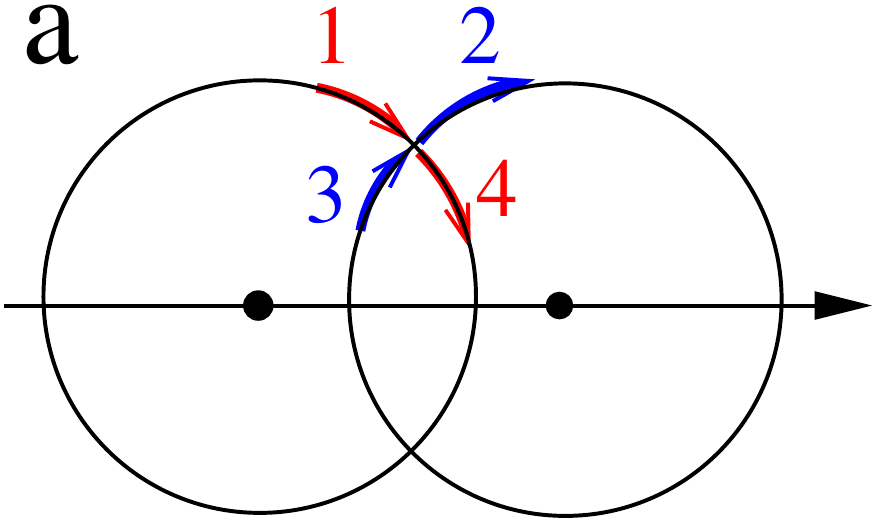}
    \hspace{20pt}
    \includegraphics[height=0.11\textwidth]{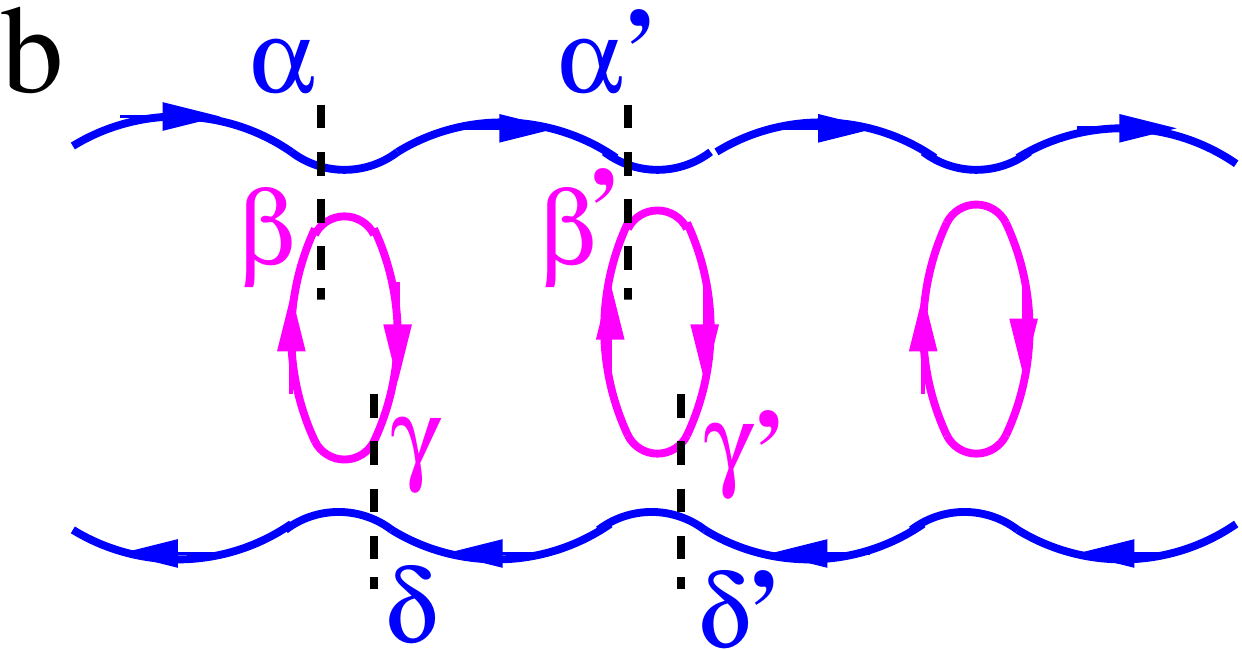}
    \caption{(a) Scattering states for a quantum mechanical jump between
    two orbits. (a) Definition of wave amplitudes at different points
    along the semi-classical trajectory. Un-primed letters are separated
    from primed letters by a momentum $(g,0)$.}
    \label{fig:scat}
\end{figure}

We put the term ``trajectory'' in inverted commas because this is not
quite a classical trajectory, there are quantum mechanical jumps at the
Brillouin zone boundaries. For a trajectory without jumps the
wave-function in the semi-classical approximation is $\psi({\bm p})
\propto e^{iS({\bm p})}$, where $S({\bm p})$ is the phase (action) along
the trajectory, though the approximation fails where two circular
trajectories intersect. An intersection point is shown in Fig.
\ref{fig:scat}a. We can describe quantum mechanical jumps by a
scattering matrix $S$, relating incoming waves 1 and 3 to ``reflected''
waves 2 and 4.

\begin{eqnarray}\label{sm}
    &&
    {\hat S}
    \left(
    \begin{array}{c}
        1 \\ 3
    \end{array}\right)\nonumber
    =
    \left(
    \begin{array}{c}
        4 \\ 2
    \end{array}\right)
    \\
    &&
    {\hat S} =
    \left(
    \begin{array}{cc}
        S_{41}& S_{43}\\
        S_{21}&  S_{23}
    \end{array}
    \right)
    =
    \left(
    \begin{array}{cc}
        s & c e^{-i \varphi} \\
        -c e^{i \varphi} & s \\
    \end{array}
    \right)
\end{eqnarray}

Since ${\hat S}$ is unitary the parameters $c$ and $s$ are real with
$c^2 + s^2 = 1$. In the perturbative limit, $W \to 0$, the wave 1
scatters to 4 and the wave 3 scatters to 2 (see Fig. \ref{fig:scat}).
Hence, in this case, we must have $s \to 1$ and $c \to 0$ (it turns out
that $\varphi \to \pi / 4$ also). In the opposite limit, the adiabatic
limit, in which $W$ is large the wave 1 scatters to 2 and the wave 3
scatters to 4. Thus, $s \to 0$, $c \to 1$ and $\varphi \to 0$. The
explicit scattering matrix is derived in the appendix in terms of the
parameters over our problem: the magnetic field, $B$, the potential
amplitude, $W$, the Fermi momentum, $p$, and the reciprocal lattice
vector $g$.

\begin{align}\label{scpar}
\begin{split}
    s & = e^{-\pi z} \\
    c & = \sqrt{ 1 - s^{2} } \\
    \varphi & =
    \frac{\pi}{4} -
    \arg ( \Gamma ( 1 + i z ) ) + z ( \ln(z) - 1 ) \\
    z & = \frac{W^2}{2 ( \omega \epsilon g / p ) \sqrt{1-g^2/4p^2}}
\end{split}
\end{align}

Here $\omega=\frac{eB}{m}$ is the cyclotron frequency and $\epsilon=p^2
/ 2 m$ is the energy of the electron. We are mostly interested in the
case $g \sim p$, so the perturbative limit corresponds to $W \ll
\sqrt{\omega \epsilon} $ and the adiabatic limit corresponds to $W\gg
\sqrt{\omega \epsilon}$. We use the term ``magnetic breakdown regime''
to refer to the intermediate case $W \sim \sqrt{\omega \epsilon}$. Of
course, we always assume that $W$ and $\omega$ are much less than
$\epsilon$. Eqns. \ref{sm} and \ref{scpar} fully describe scattering around 
the intersection of two circular orbits.

Next, we account for the periodicity of Fig.
\ref{fig:gratingFermiSurfaceA0A1}. In Fig. \ref{fig:scat}b we denote by
Greek letters the wave functions at the point on the trajectory just
before scattering. The primed points are separated from the un-primed
points by one period, $g$. According to Bloch's theorem the wave
functions are related by

\begin{align}\label{BL}
\begin{split}
    & \alpha' = e^{ik} \alpha \\
    & \beta'  = e^{ik} \beta  \\
    & \gamma' = e^{ik} \gamma \\
    & \delta' = e^{ik} \delta
\end{split}
\end{align}

Where $e^{ik}$ is a phase factor and $k$ is the ``quasi-momentum". Free,
semi-classical propagation away from the intersections is described by
the matrix

\begin{eqnarray}\label{freprop}
    {\hat P} =
    \left(
        \begin{array}{cc}
            P_m & 0 \\
            0 & P_b
        \end{array}
    \right)
\end{eqnarray}

Here $P_b = e^{iS_b}$ describes propagation along the blue line in Fig.
\ref{fig:scat}b from $\alpha$ to $\alpha^{\prime}$ or from
$\delta^{\prime}$ to $\delta$. And $P_m = e^{iS_m - i\pi/2}$ describes
propagation along half of the magenta oval in Fig. \ref{fig:scat}b. Each
half of this oval contains a classical turning point, leading to the
caustic phase $\pi / 2$ in $P_{m}$. The actions $S_b$ and $S_m$ are
related to the areas $A_0$ and $A_1$ introduced above as

\begin{align}\label{Sbm}
\begin{split}
    &
    S_m = \frac{A_1}{2|eB|} \\
    &
    S_b + S_m = \frac{A_0}{2|eB|}
\end{split}
\end{align}

Combining scattering events and propagations gives a relation between
the amplitudes defined in Fig. \ref{fig:scat}

\begin{align}\label{prop1}
\begin{split}
    &
    \left(
        \begin{array}{c}
            \gamma \\
            \alpha'
        \end{array}
    \right) =
    {\hat P}{\hat S}
    \left(
        \begin{array}{c}
            \alpha \\
            \beta
        \end{array}
    \right) \\
    &
    \left(
        \begin{array}{c}
            \beta' \\
            \delta
        \end{array}
    \right) =
    {\hat P}{\hat S}
    \left(
        \begin{array}{c}
            \delta' \\
            \gamma'
        \end{array}
    \right)
\end{split}
\end{align}

Equations \ref{BL} and \ref{prop1} are a system of four equations with
four coefficients: $\alpha$, $\beta$, $\gamma$, $\delta$. The right hand
side of this system of equations is zero meaning that non-trivial
solutions occur only when the determinant is equal to zero. In a
simplified form this condition reads
\begin{align}\label{eqn:quantisationCondition}
\begin{split}
    0 =
    \cos \left( \frac{A_{0}}{2|eB|} \right) +
    &
    2 c \cos (k)
    \cos \left( \frac{A_{1}}{2|eB|} - \varphi \right) + \\
    &
    c^{2} \cos \left( \frac{A_{0} - 2 A_{1}}{2|eB|} + 2 \varphi \right)
\end{split}
\end{align}
The areas $A_{0}$ and $A_{1}$ can be found explicitly in the limit $W
\to 0$.
\begin{align}\label{areas}
\begin{split}
    &
    A_0 = \pi p^2 \\
    &
    A_1 = 2 p^2 \arcsin\sqrt{1-\frac{g^2}{4p^2}} -
    gp \sqrt{1-\frac{g^2}{4p^2}}
\end{split}
\end{align}
These areas, as well as the parameters of the scattering matrix (Eq.
\ref{scpar}), depend on the energy, $\varepsilon = p^{2} / 2 m$. Eq.
\ref{eqn:quantisationCondition} is therefore a highly nonlinear function
of energy which allows us to find the energy levels,
$\varepsilon_{n,k}$, of the system at a given magnetic field, $B$. The
energy levels depend on an integer quantum number $n$ that originates
from the Landau level index and on the continuous parameter $k$ limited
by $- \pi < k < + \pi$. Eq. \ref{eqn:quantisationCondition} can be
solved analytically in certain simple cases; that is, when $c = 0$ or $c
= 1$.

The approach described above was developed by Pippard in Ref.
\cite{pippard_quantization_1962}. However, the Pippard work derives Eq.
\ref{eqn:quantisationCondition} without account of the reflection phase,
$\varphi$.

\subsection{Density of states in the weak scattering limit.}

The density of states, $\rho$, is the simplest equilibrium property
containing magnetic oscillations. We present an explicit calculation of
the oscillating components of $\rho$ in this section as an instructive
example which leads on to our later calculations of the thermodynamic
potential. Our expression for $\rho$ contains some of the essential
features of the problem: the lack of non-Onsager frequencies in the
single particle picture. Importantly, this calculation cannot address
the temperature dependence of magnetic oscillations; that requires a
study of the full thermodynamic potential.

We consider the quantization condition (Eq.
\ref{eqn:quantisationCondition}) analytically in the limit of weak
scattering, $W^2 \ll \omega \epsilon$, which is realised in the
experiment Ref. \cite{wang_formation_2023}. In this limit $c$ is a small
parameter, so we can solve Eq. \ref{eqn:quantisationCondition}
perturbatively, in powers of $c$. Here, as well as in the rest of analytical
calculations in this paper, we work to second-order in $c$.
First, the zeroth order solution ($c = 0$) of Eq.
\ref{eqn:quantisationCondition} gives usual Landau levels.
\begin{eqnarray}\label{Ll}
    \epsilon_{n}^{(0)}=\omega(n+1/2)
\end{eqnarray}
Where $n = 0, \ 1, \ 2, \ \cdots$. The degeneracy of each Landau level
arises from the free parameter $k$. The solution of Eq.
\ref{eqn:quantisationCondition} to second order in $c$ is
straightforward, it depends on the energy derivatives, $A' = \partial A
/ \partial \varepsilon$, and on the areas, $A$, each of which must be
evaluated at $\varepsilon = \varepsilon_{n}^{(0)}$. In particular,
$A_{0} = 2 \pi m \omega (n + 1/2)$ and $A_{0}' = 2 \pi m$. To
second-order in $c$ the parameters of the scattering matrix (Eq.
\ref{scpar}) are
\begin{align}\label{smallc}
\begin{split}
    s & \approx 1 - c^2 / 2 \\
    \varphi & \approx 
    \frac{\pi}{4} -
    \frac{c^2}{2\pi}
    \left( \ln \frac{2\pi}{c^2} + 0.42 \right)
\end{split}
\end{align}
And the solution of Eq. \ref{eqn:quantisationCondition} for the energy
levels, $\varepsilon_{n,k}$, is
\begin{align}\label{eqn:pippardSolution}
\begin{split}
    \epsilon_{n,k} = &
    \epsilon_{n}^{(0)} + \delta \epsilon_{n,k} \\
    \delta\epsilon_{n,k} = &
    \frac{2 \omega}{\pi} c \cos(k)
    \sin \left( \frac{A_0-A_1}{2|eB|} + \varphi \right) + \\ &
    \frac{\omega}{\pi} c^{2}
    \sin\left(\frac{A_1}{|eB|}- 2 \varphi\right)
    \left[ 1 - 2\frac{A_{1}'}{A_{0}'} \cos^{2}k \right]
\end{split}
\end{align}
The degeneracy of each Landau level is lifted due to the quasi-momentum, $k$. 

We can obtain the density of states from this expression using
\begin{eqnarray}\label{ds1}
    \rho(\epsilon,B) =
    \sum_n \int_{-\pi}^{\pi}
    \delta( \epsilon - \epsilon_{n,k} ) \frac{dk}{2\pi}
\end{eqnarray}
The above solution implies that the magnetic field is fixed and energy
is variable. We can also fix the energy equal to the chemical potential,
$\epsilon = \mu = \epsilon_F$, and vary the magnetic field. This
representation is more convenient for magnetic oscillations. In this
case the areas $A_0 = A_0(\mu)$ and $A_1 = A_1(\mu)$ in
Eq. \ref{eqn:quantisationCondition} are fixed, but magnetic field
varies. The density of states at the Fermi energy is then
\begin{align*}
    \rho(B ) \propto \sum_{n, k} \delta( B - B_{n,k})
\end{align*}
Where $B_{n,k}$ are solutions to Eq. \ref{eqn:quantisationCondition}.
Hereafter we assume that $B>0$ and omit the absolute value in $|eB|$.
Repeating the second-order expansion of Eq.
\ref{eqn:quantisationCondition} in this representation gives
\begin{eqnarray}
\label{bnklsol}
    \frac{A_0}{eB_{n, k}} &= &
    2 \pi ( n + 1 / 2 ) \\
    & +&
    4 c \cos(k)
        \sin \left[ \pi (1 - \gamma) \left( n + 1/2 \right) +
                    \varphi
             \right]\nonumber \\ 
             & +&(
    2 c^{2} ( 1 - 2 \gamma \cos^{2} (k))
        \sin \left[ 2 \pi \gamma \left( n + 1/2 \right) -
                    2 \varphi
             \right]\nonumber\\
 \gamma &=& \frac{A_1}{A_0}\nonumber             
\end{eqnarray}

Magnetic oscillations are periodic in $1/B$, we address
oscillations by taking the Fourier transform of the density of states
with respect to $1/B$.
\begin{align}\label{eqn:dosFourierGeneral}
    \rho_q = \int e^{i q/B}\rho(B)d(1/B)
    \propto \sum_{n,k} e^{i q/B_{n, k}}
\end{align}
We substitute $1/B_{n,k}$ from Eq. \ref{bnklsol} to obtain
\begin{align}\label{Ft1}
\begin{split}
    \rho_{q} \propto
    \sum_n
    & e^{i 2 \pi Q (n + 1/2 )} \times
    \\ &
    \left[
        ( 1 - 2 c^{2} Q^2 )
    \right.
    \\ &
        - c^2 Q (1 - \gamma + Q)
        e^{- i 2 \pi \gamma ( n + 1/2 ) + 2 i \varphi}
    \\ &
        + c^2 Q (1 - \gamma - Q)
    \left.
        e^{+ i 2 \pi \gamma ( n + 1/2 ) - 2 i \varphi}
    \right]
\end{split}
\end{align}
Where
\begin{align*}
    Q = \frac{q}{|e| A_0}
\end{align*}
We remind that the convention is for frequency to be positive, $Q > 0$.
Eqn. \ref{Ft1} contains several harmonics:
\begin{enumerate}

    \item The second line in Eq. \ref{Ft1} gives a peak in the Fourier
    transform at integer $Q$, leading to frequencies $q = e A_{0} n$ for
    $n = 1, \ 2, \ \cdots$. This is the usual Onsager frequency for
    $A_{0}$; its amplitude, $1 - 2 c^{2} n^{2}$, is non-zero even at $c
    = 0$.

    \item The third line in Eq. \ref{Ft1} gives a Fourier peak when $Q
    - \gamma = n$ for $n = 0, \ 1, \ 2, \ \cdots$. In terms of
    the frequency we have $q = e A_{0} n + e A_{1}$. In particular, this
    harmonic leads to $q = e A_{1}$ and to $q = e ( A_{0} + A_{1} )$.
    The corresponding amplitudes are proportional to $c^2$.

    \item The fourth line in Eq. \ref{Ft1} gives a Fourier peak when $Q
    + \gamma = n$ for $n = 0, \ 1, \ 2, \ \cdots$ or $q = e
    A_{0} n - e A_{1}$. From this harmonic we obtain the Onsager
    frequency $q = e (2 A_{0} - A_{1})$. Notably, this condition also
    allows for the non-Onsager frequency $q = e(A_{0} - A_{1})$ when $Q
    + \gamma = 1$. However, computing the amplitude of this harmonic,
    $c^{2} Q (1 - Q - \gamma)$, gives exactly zero.
\end{enumerate}
Thus, the density of states oscillates with a variety of different
Onsager frequencies. Non-Onsager frequencies are notably not present.
\begin{figure}[ht]
    \centering
    \includegraphics[width=0.4\textwidth]{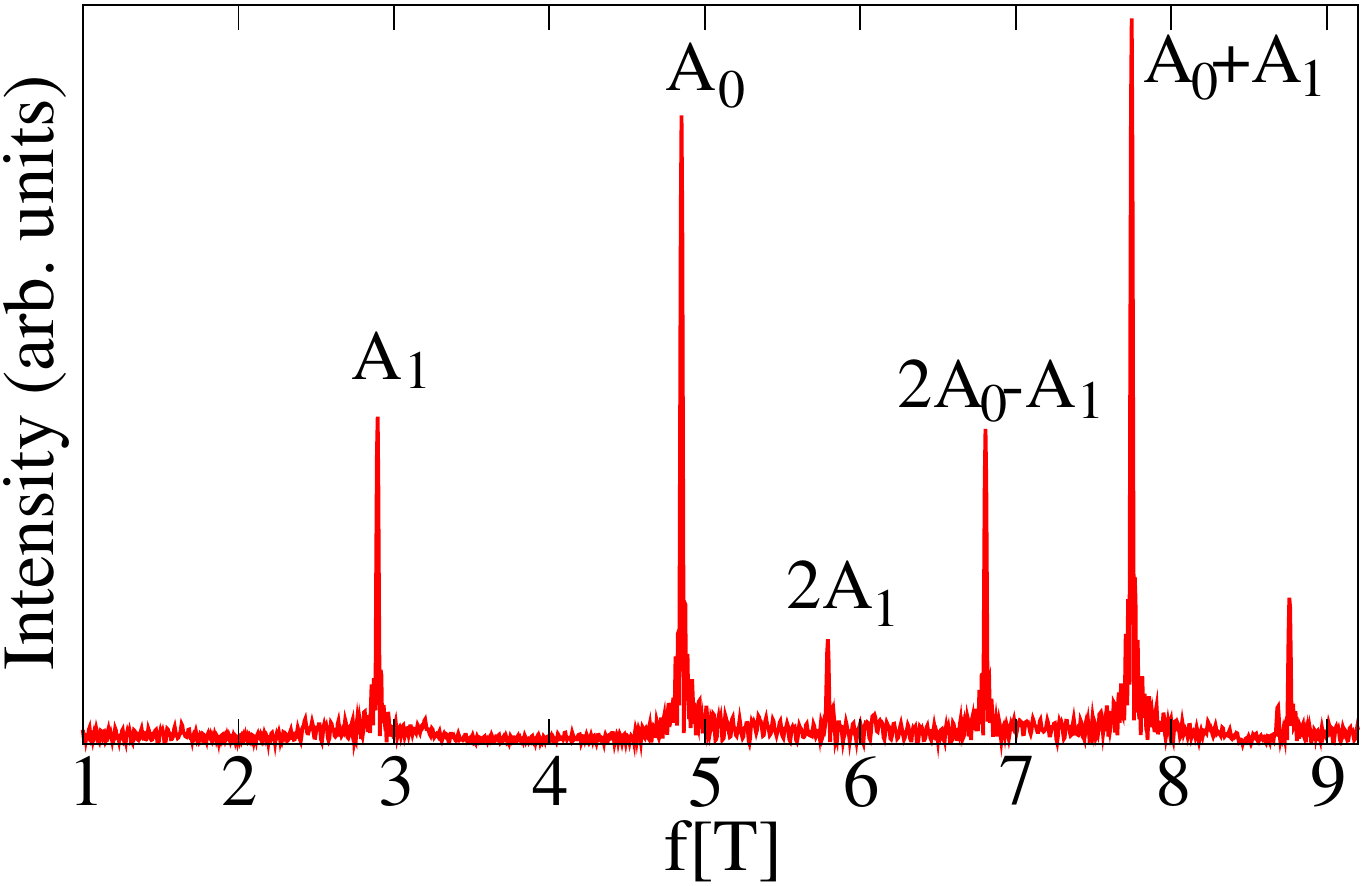}
    \caption{Fourier transform of the density of states vs frequency
        obtained by numerical solution of Eq.~\ref{eqn:quantisationCondition}.
There are peaks at $A_0$ ($f=4.85$\,T), $A_1$, and at  combinatorial frequencies,
but there are no peaks at non-Onsager frequencies.
The parameters are $\epsilon_F=8$\,meV, $a=80$\,nm, $c=0.6$.
}       
    \label{fig:Pip}
\end{figure}

The above analytical analysis is valid up to the second order in $c$.
To verify our conclusion in any order in $c$ we solve Eq.~\ref{eqn:quantisationCondition} exactly 
numerically. Here we present results  for GaAs with $m=0.07m_e$, the Fermi energy $\epsilon_F=8$\,meV 
(corresponds to the density $n=2.36\times 10^{11}\,\textrm{cm}^{-2}$), the superlattice spacing $a=80$\,nm, and the tunneling
matrix element $c=0.6$.
The Fourier transform of the DOS versus frequency $f$  is plotted in Fig.~\ref{fig:Pip}.
The frequency has dimension Tesla and is defined as $q=2\pi f$, see Eq.~\ref{eqn:dosFourierGeneral}.
The Fourier transform shows peaks at $A_0$ ($f=4.85$\,T), $A_1$, and various combinatorial frequencies,
however, there are no peaks at non-Onsager frequencies.

\section{Exact numerical solution of the single particle Schr\"odinger equation}

In the present section we solve exactly numerically the single particle Schr\"odinger equation
using the KWANT software package \cite{Groth2014}. The important point is that the package
allows us to calculate the density of states and multi-terminal resistances, meaning it can
address both equilibrium and non-equilibrium properties of the system. 
The aim of the present section is twofold: first, to check by another method (KWANT) that
$\rho$ does not contain non-Onsager oscillations, and second, to check that the 
resistance---a non-equillibrium property---contains non-Onsager oscillations and hence to
demonstrate a correspondence with previous works based on the semi-classical kinetic equation 
\cite{morrison_two-lifetime_1981, beenakker_guiding-center-drift_1989, winkler_landau_1989,
mirlin_weiss_1998}.

\begin{figure}[t]
    \includegraphics[width=0.3\textwidth]{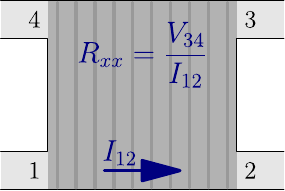}
    \caption{Schematic of the four-terminal device simulated using KWANT. One-dimensional potential modulation along $x$ is shown by vertical lines.}
    \label{scheme}
\end{figure}

We solved the quantum mechanical problem of electron scattering for 
a two-dimensional system with either a one-dimensional or a two-dimensional potential modulation. 
A schematic of the device is shown in Fig.\ref{scheme}. The lattice potential 
is defined in the central square, on each side of which 
there are two horizontal channels. Electrons incident on the lattice from one channel are
scattered into the three other channels. Calculations were performed for
devices containing 40 to 50 lattice periods. In the schematic (Fig. \ref{scheme}), 
the vertical lines show a one-dimensional modulation of potential along $x$, 
with the current between contacts 1 and 2 directed perpendicular to 
the equipotential lines. The magnetic field is directed perpendicular to the 2D plane.

KWANT calculates the density of states and four-terminal transmission 
coefficients for given values of energy $\epsilon_F$ and magnetic field $B$.
Four-terminal resistances are determined from the calculated transmission 
coefficients within the framework of the Landauer-B\"uttiker formalism~\cite{Buttiker1986}. 
We then apply a Fourier transform to the calculated $\rho(1/B)$ and 
$R_{xx}(1/B)$ in order to determine the characteristic frequencies.

We consider first the one-dimensional potential described by Eqn. \ref{u1D}. In the case
of an 80\,nm period the potential is defined on a square mesh with the 
mesh spacing 8\,nm. To remove interference fluctuations related to scattering from 
the device boundaries we introduce a random disorder $-V_r/2 < V < V_r/2$ 
on every site of the mesh \cite{Tkachenko2022, Tkachenko2023}.

\begin{figure}[ht]
    \includegraphics[width=0.45\textwidth]{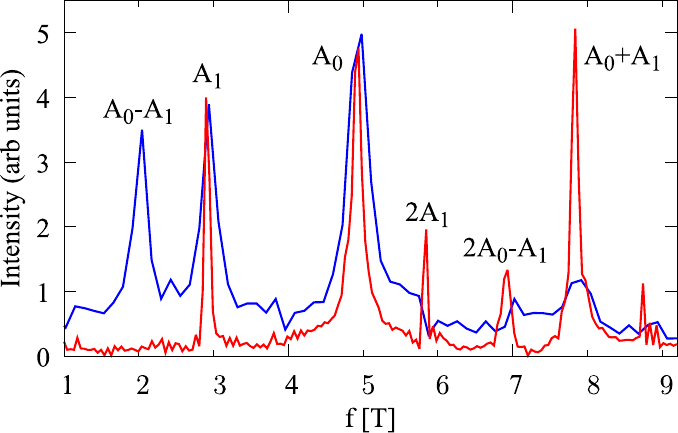}
    \caption{Fourier transforms of $\rho$ (red) and $R_{xx}$ (blue) vs frequency for a one-dimensional superlattice. The results are obtained  using KWANT.  The parameters are $\epsilon_F=8$\,meV, $a=80$\,nm, $W=0.4$\,meV,  $V_r=0.5$\,meV. 
    }
    \label{fig-fft-kwant}
\end{figure}

In Fig.~\ref{fig-fft-kwant} we present the results of our KWANT calculation for the same
parameters used in Fig.~\ref{fig:Pip}: the Fermi energy $\epsilon_F=8$\,meV 
(corresponding to a density $n=2.36\times 10^{11}\,\textrm{cm}^{-2}$), the superlattice 
period $a=80$\,nm, and the superlattice amplitude $W=0.4$\,meV.
The density of states is given by the red line in Fig. \ref{fig-fft-kwant} and shows perfect agreement with the results of the semi-classical calculation (Fig. \ref{fig:Pip}). In 
particular the density of states contains only Onsager frequencies.
The resistance, however, manifests the non-Onsager frequency $A_{0} - A_{1}$ in addition to other Onsager frequencies (blue line in Fig. \ref{fig-fft-kwant}).

The presence of the non-Onsager line $A_0-A_1$ in the resistance is consistent
with previous results using kinetic equation theory \cite{morrison_two-lifetime_1981,
beenakker_guiding-center-drift_1989, winkler_landau_1989,
mirlin_weiss_1998}, however, there is a difference.
According to the semi-classical theory \cite{
beenakker_guiding-center-drift_1989, winkler_landau_1989,
mirlin_weiss_1998} the period of oscillations is determined by the condition
that the diameter of the cyclotron orbit is equal to an integer number of
lattice periods, this is why quite often these oscillations are
called "commensurability oscillations". This condition gives the frequency of oscillations
\begin{eqnarray}
    \label{commensur}
 f= \frac{gp_{F}}{\pi}  
\end{eqnarray}
According to our calculation the frequency is
\begin{eqnarray}
    \label{exact}
 f= \frac{A_0-A_1}{2\pi} \ ,  
\end{eqnarray}
where $A_0$ and $A_1$ are given by Eqn. \ref{areas}. While in the classical limit ($g \ll p_{F}$)
Eqns. \ref{commensur} and \ref{exact} are identical, at $g\to 2p_{F}$ the 
approximate commensurability expression (Eqn. \ref{commensur})  differs from the exact one (Eqn. \ref{exact})
by 27\%.

\begin{figure}[t]
    \includegraphics[width=0.45\textwidth]{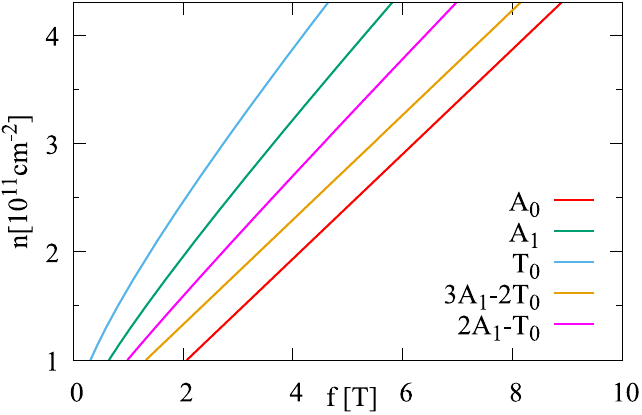}
    \caption{Fourier peaks of density of states $\rho$ for a 2D
    triangular modulation. Lines indicate positions of the Fourier
    transform peaks on the density-frequency plane. The potential
    amplitude is $W=0.3$\,meV, the amplitude of disorder is
    $V_r=1$\,meV. Only Onsager frequencies are observed in $\rho$. }
    \label{fig2}
\end{figure}

\begin{figure}[t]
    \includegraphics[width=0.15\textwidth]{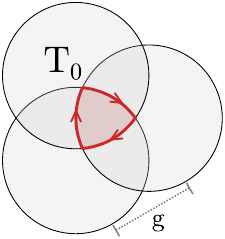}
    \caption{Triangular electron trajectory $T_{0}$ --due to magnetic breakdown in a triangular periodic potential (Eqn. \ref{u2D}).}
\label{fig:trajectories}
\end{figure}

We have also used KWANT to study a two-dimensional triangular potential
\begin{align}\label{u2D}
    U({\bm r}) =
    2 W \left[ \cos({\bm g}_1\cdot {\bm r}) +
               \cos({\bm g}_2\cdot {\bm r}) +
               \cos({\bm g}_3\cdot {\bm r})
        \right]
\end{align}
Here $\mathbf{g_i}$ are the basic vectors of the honeycomb Brillouin zone. 
We use the same lattice spacing as for the 1D modulation, $a=80$\,nm, and the same as that in
the experimental work Ref. \cite{wang_formation_2023}.
Fig. \ref{fig2} presents our results for $\rho$: positions of the Fourier transform peaks
on the density-frequency plane. Since there are a larger number of possible orbits on the triangular lattice, we present the dependence of frequency on density for each Fourier peak, which allows us to pin down exactly which orbit corresponds to which line. Again, only Onsager frequencies are observed in $\rho$,
the corresponding trajectories are marked for every line,
some of these trajectories are related to the triangular orbit $T_{0}$ shown in
Fig.\ref{fig:trajectories}.

\begin{figure}[t]
    \includegraphics[width=0.45\textwidth]{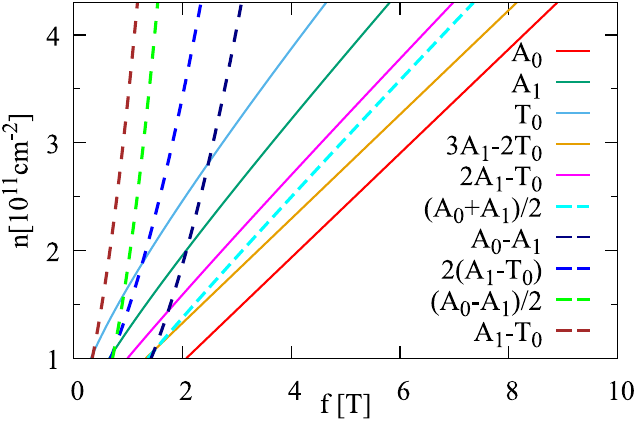}
    \caption{ 
    Nonlocal resistance $R_{xx}$ for 2D triangular modulation:  positions of the Fourier transform peaks 
    on the density-frequency plane. The potential amplitude $W=0.3$\,meV, the amplitude of disorder $V_r=1$\,meV. Onsager lines are solid and non-Onsager ones are dashed.
 }
    \label{fig3}
\end{figure}

Fig. \ref{fig3} presents our results for resistance: similar to Fig. \ref{fig2} we plot the central frequency of each Fourier peak as a function of density.
Similar to what we observed for a 1D superlattice the resistance manifests non-Onsager 
frequencies (several of them, including $A_{0} - A_{1}$), moreover, it
even manifests half-frequencies (e.g. $(A_{0} - A_{1})/2$). These non-Onsager frequencies are indicated by the dashed lines in Fig. \ref{fig3}; the solid lines show Onsager frequencies that were also observed in the density of states.

The numerical results of this section explain all existing transport experiments,
including the most recent one, Ref. \cite{wang_formation_2023}. In other words, our numerics
fully explain the Shubnikov–de Haas effect in the magnetic breakdown regime. 
However, as we pointed out in the introduction, non-Onsager
oscillations have been also observed in the equilibrium de Haas–Van Alphen effect 
\cite{eddy_haas---van_1982,audouard_quantum_2012} and any kinetic mechanisms are irrelevant
in this case.
The rest of this paper is concerned with the equilibrium case. To address this case we need to 
consider the thermodynamic $\Omega$-potential.

\section{Thermodynamic potential of non interacting (ideal) electron gas}\label{sec:singleParticleOmega}

Thus far we have shown that non-Onsager frequencies \emph{cannot} appear
in the density of states. We turn now to our result related to the interaction:
non-Onsager frequencies arise due to Coulomb electron-electron interactions.
This is the only mechanism that contributes to the de Haas–Van Alphen effect, of
course the mechanism also contributes to  the  Shubnikov–de Haas effect.
Specifically, it will be shown
that an $A_{0} - A_{1}$ frequency appears in the thermodynamic
potential, $\Omega$, with the correct temperature dependence.

We present two calculations; instead of dealing with magnetisation
directly we deal with $\Omega$. First, in this section, we calculate
$\Omega$ without Coulomb interactions. This result reiterates the
results of the previous section and introduces our basic technique.
Second, in the following section, we calculate the correction $\delta
\Omega$ to the thermodynamic potential due to a repulsive contact
interaction between electrons. This correction contains non-Onsager
frequencies.

The thermodynamic potential has of course already been computed in a
general form by Lifshitz and Kosevich\cite{lifshits_theory_1956}, however, 
the calculation does not include both magnetic breakdown and Coulomb interactions. To account
for magnetic breakdown we make use of the energies Eq. (\ref{eqn:quantisationCondition})
obtained within the Pippard model. Coulomb interactions are accounted
for in the following section using a technique which builds on the one
developed here.

The $\Omega$-potential of an ideal gas with a set of energy levels
$\varepsilon_{n, k}$ is given by
\begin{align}\label{Omega1}
\begin{split}
    \Omega & =
    -T \frac{eB}{\pi}
    \sum_{n,k} \ln \left( e^{(\mu-\epsilon_{nk})/T} + 1 \right) \\
    & \propto \int d\epsilon \ \rho(\epsilon,B)
    \ln \left( e^{(\mu-\epsilon)/T} + 1 \right)
\end{split}
\end{align}
This is a linear functional of the density of states,
$\rho(\epsilon,B)$; since the non-Onsager frequencies are absent from
$\rho$ they must also be absent from $\Omega$. Note that we normalize
the k-integration as $\sum_k=1$. The contribution of each Landau level
must therefore be multiplied by the capacity of the level $eB/\pi$.

We compute $\Omega$ up to second order in the breakdown amplitude $c$
using Eqn.(\ref{eqn:pippardSolution}): $\epsilon_{nk} = \epsilon_n^{(0)}
+ \delta\epsilon_{nk}$. The calculation consists of 4 steps. (i) First
expand the logarithm in the 1st line of Eq. (\ref{Omega1}) up to the
second order in $\delta \epsilon_{nk}$. Starting from $F(\epsilon_{nk})
= \ln (e^{(\mu-\epsilon_{nk})/T} + 1 )$ we obtain
\begin{align}\label{FF}
    F(\epsilon_{nk}) = F(\epsilon_n^{(0)}) +
    \frac{\partial F}{\partial \epsilon} \delta \epsilon_{nk} +
    \frac{1}{2} \frac{\partial^2 F}{\partial
    \epsilon^2}\delta\epsilon_{nk}^2
\end{align}
After this expansion the integration over $k$ is performed, eliminating
the linear in $c$ term (see Eqn. \ref{eqn:pippardSolution}). (ii) Next,
in each term of Eqn. \ref{FF} we replace summation over n by integration
over $x$ using the Poisson formula.
\begin{align}\label{Ps}
\begin{split}
    \sum_{n=0}^{\infty} \bar{F}(n) =
    \frac{1}{2} \bar{F}(0) + &
    \int_0^{\infty} \bar{F}(x) dx + \\ &
    2 Re \sum_{l=1}^{\infty}
    \int_0^{\infty}
    \bar{F}(x) e^{2\pi i l x} dx
\end{split}
\end{align}
Here $\bar{F}(n)$ represents $\sum_{k} F(\varepsilon_{n,k})$. 
So, the discrete index n is replaced by the continuous variable x.
(iii) In the 3rd step we perform, if necessary, the x-integration by parts to bring
the Fermi-Dirac distribution to the bell-shape function peaked at the
chemical potential.
\begin{align*}
    \frac{e^{(\epsilon_x^{(0)}-\mu)/T}}
         {\left(e^{(\epsilon_x^{(0)}-\mu)/T}+1\right)^2}
\end{align*}
(iv) In the final step we perform the integration over $x$ using
\begin{align*}
    \int_{-\infty}^{+\infty} e^{i \alpha x}
    \frac{e^x}{(e^x+1)^2} =
    \frac{\pi \alpha}{\text{sh}(\pi \alpha)}
\end{align*}
This calculation results in the following $\Omega$-potential.
\begin{align}\label{Oid}
\begin{split}
    \Omega =
    -\frac{m}{2\pi} \mu^2
    - \frac{m\omega^2}{2\pi^3} &
    \left[
        \frac{t_0}{\text{sh}(t_0)}
        \cos \left( \frac{A_0}{eB} \right) + \right.
        \\ & \left.
        \frac{c^2}{\alpha}
        \frac{t_1}{\text{sh}(t_1)}
        \cos \left( \frac{A_1}{eB}-2\varphi \right)
    \right]
\end{split}
\end{align}
Where
\begin{align*}
\begin{split}
    t_0    = & \frac{d A_{0}}{d \varepsilon} \frac{\pi T}{e B}
           = {2\pi^2T}/{\omega} \\
    t_1    = & \frac{d A_{1}}{d \varepsilon} \frac{\pi T}{e B}
           = \alpha{2\pi^2T}/{\omega} \\
    \alpha = & \frac{dA_1}{d\epsilon}
               \left/ \frac{dA_0}{d\epsilon} \right.
\end{split}
 \end{align*}
The first term in the oscillating part of the $\Omega$-potential is the
usual Lifshitz-Kosevich expression, and the second term corresponds to
the $A_1$-frequency. The temperature dependence of the $A_1$-term is
different from that of the $A_0$-term. As we explained the non-Onsager
frequency $A_0-A_1$ does not appear. Of course, there are terms with
other Onsager frequencies, we just do not present them.

\section{Thermodynamics potential with electron-electron interactions}

One of our main results is that the inclusion of electron-electron
interactions leads to a finite amplitude for non-Onsager frequencies
with the correct temperature dependence. We stress that this is an
entirely equilibrium effect. While the calculation is somewhat involved
the procedure is straightforward: we substitute the energy levels
obtained from the Pippard model into the leading interaction correction
to the $\Omega$-potential and then expand to second order in the magnetic
breakdown parameter, $c$.

With account of screening the electron-electron Coulomb interaction in
momentum space reads
\begin{eqnarray}
\label{SCR}
\frac{2\pi e^2}{q-2\pi e^2P_q}\to \frac{2\pi e^2}{q+2m e^2}\to \frac{\pi}{m} 
\end{eqnarray}
Here we take into account that at a relevant momentum transfer, $q < 2p_F$, the polarization operator is $P_q=-m/\pi$, and also neglect $q$ compared to $2me^2$. 
Transferring (\ref{SCR}) to the coordinate representation we get
the repulsive contact interaction
\begin{eqnarray}\label{int1}
    H_{int} &=& \lambda \delta( {\vec r}_1 - {\vec r}_2 )\nonumber\\
    \lambda &=& \frac{\pi}{m}
\end{eqnarray}

\begin{figure}[h]
    \caption{Diagram for the leading-order interaction correction to
    $\Omega$. Solid lines represent electrons. The dashed line
    represents the Coulomb interaction.}
    \includegraphics[width=0.15\textwidth]{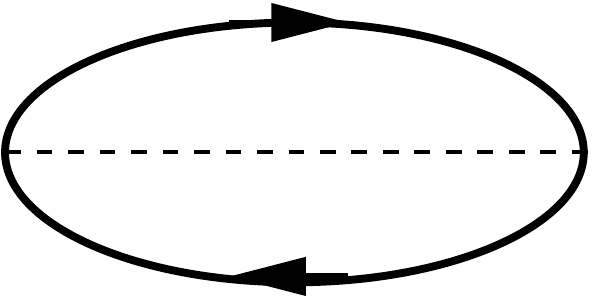}
    \label{Fexc}
\end{figure}

The leading-order interaction correction to the $\Omega$-potential is given by
the exchange diagram in Fig. \ref{Fexc}.

The correction shown in Fig. \ref{Fexc} is given explicitly by\cite{landau_chapter_1980}

\begin{align}\label{int2}
\begin{split}
    \delta \Omega = &
    -\frac{1}{2} \sum_{ij}
    \langle
        \psi_j^*(r_1) \psi_i^*(r_2) 
        |H_{int|}|
        \psi_i(r_1)\psi_j(r_2)
    \rangle
    f_i f_j \\
    f_i = & \frac{1}{ e^{(\epsilon_i - \mu)/T} + 1 }
\end{split}
\end{align}
Here, the indices $i$ and $j$ enumerate quantum states; the index
describes both orbital motion and spin. In what follows we skip spin and
accordingly skip the pre-factor 1/2 in (\ref{int2}). Without magnetic
field and at $T=0$ the correction can be immediately calculated using
plane-wave quantum states, $\psi_i \propto e^{i \bm{p}_i \cdot \bm{r}}$.
\begin{eqnarray}\label{int3}
    \delta \Omega(B = 0, T = 0) =
    -\frac{1}{4} \lambda n^2 =
    -\frac{m}{4\pi} \mu^2
\end{eqnarray}
This correction is just two times smaller that the first term in Eq.
(\ref{Oid}). We are interested in non-zero magnetic fields, however, and
in this case the wave-functions depend on gauge. We work with vector
potential $A = (0, Bx, 0)$, so the single electron wave-function is
\begin{eqnarray}\label{wf1}
    \psi_{nk} = e^{iky} \chi_n(x+k/eB)
\end{eqnarray}
Where $\chi_n$ is the oscillator wave function corresponding to the
Landau level $n$. We reiterate that in the 1D model (Eq. \ref{u1D}) the
potential is modulated along the $x$-direction so that $k$ in Eq.
\ref{wf1} remains a good quantum number in the presence of the periodic
potential. Note that the $k$ which arises in the wave-function (Eqn. \ref{wf1}) differs from the dimensionless $k$ in the semi-classical expression for energy levels (Eqn. \ref{eqn:pippardSolution}). In terms of the $k$ from Eqn. \ref{wf1}, the dimensionless parameter in Eqn. \ref{eqn:pippardSolution} is given by $g k / e B$.

Now, we can rewrite the exchange correction to $\Omega$ (Eqn.
\ref{int2}) as
\begin{eqnarray}\label{int7}
    \delta \Omega =
    -\frac{eB}{2\pi}
    \sum_{n_1, k_1,n_2, k_2}
    V_{n_1, k_1,n_2, k_2} f_{n_1, k_1} f_{n_2, k_2}
\end{eqnarray}
Where $V$ is the exchange matrix element of the interaction and is given
by
\begin{align}\label{int4}
\begin{split}
    V_{n_1, k_1, n_2, k_2} & = \lambda \int dx
    \chi_{n1}^2 (x)
    \chi_{n2}^2 (x+q/eB) \\
    q & = k_1 - k_2
\end{split}
\end{align}
The pre-factor in Eq. \ref{int7} is fixed by evaluating this expression
at $T = 0$ with neglect of $\delta \varepsilon_{nk}$ and then
comparing with the result of Eq. \ref{int3} (found using plane-waves).

%
%

In the following calculation it is convenient to define the average
matrix element $V^{(0)}$
\begin{align}\label{int5}
\begin{split}
    V_{n_1,n_2}^{(0)}
    & = \sum_q V_{n_1,k_1,n_2,k_2} \\
    & = \lambda
    \int dx \chi_{n1}^2(x)
    \chi_{n2}^2(x+x_q) \frac{dq}{2\pi} \\
    & = \lambda \frac{eB}{2\pi} \ .
\end{split}
\end{align}
Here $x_q=\frac{q}{eB}$. Similarly we define  $V^{(1)}$,
\begin{align}\label{int6}
\begin{split}
    V_{n_1,n_2}^{(1)}
    & = \sum_q V_{n_1, k_1, n_2, k_2} \cos(x_{q} g) \\
    & = \lambda \int dx \frac{dq}{2\pi}
    \chi_{n_{1}}^{2}(x)
    \chi_{n_{2}}^{2}(x + x_{q})
    \cos(x_{q} g) \\
    & = \lambda \frac{e B}{2\pi} M_{n_{1}} M_{n_{2}} \\
    M_{n} &=
    L_{n}( l_{B}^{2} g^{2} / 2 )
    e^{-l_{B}^{2} g^{2} / 4}
\end{split}
\end{align}
Here $g = 2 \pi / a$ is the reciprocal vector of the potential
modulation (Eq. \ref{u1D}), $l_B=1/\sqrt{eB}$ is the magnetic length
and $L_{n}$ is a Laguerre polynomial. Note
that in the limit of large $n$ we have $M_{n} \rightarrow J_{0}(\sqrt{2n} l_{B} g)$.

Next we compute the oscillating part of $\delta \Omega$. Following the
procedure used in the previous section for the ideal gas, we expand
$\delta \Omega$ (Eq. \ref{int7}) up to second order in $c$ using the
energy levels $\varepsilon_{n,k} = \varepsilon_{n}^{(0)} + \delta
\varepsilon_{n,k}$ (see Eq. \ref{eqn:pippardSolution}). Expanding the
Fermi-Dirac distribution up to second order in $\delta \epsilon_{n,k}$
gives
\begin{eqnarray}\label{ff1}
    f(\epsilon_{nk}) = f(\epsilon_n^{(0)}) +
    f'(\varepsilon_{n}^{(0)})
    \delta \epsilon_{n,k} +
    \frac{1}{2}
    f''(\varepsilon_{n}^{(0)})
    \delta \epsilon_{n,k}^{2}
\end{eqnarray}
This expression can then be substituted into Eqn. \ref{int7} for $\delta
\Omega$. Performing the integration over $k$ introduces terms involving
$\bar{f}_{n} \equiv \sum_{k} f(\varepsilon_{n,k})$ and transforms Eq.
\ref{int7} into
\begin{widetext}
\begin{align}\label{fact}
\begin{split}
    \delta \Omega = &
    - \frac{eB}{2\pi}
    \sum_{n_{1}, n_{2}}
    \left[
        V_{n_1, n_2}^{(0)}
        \bar{f}_{n_{1}}
        \bar{f}_{n_{2}} +
        V_{n_1n_2}^{(1)}
        \frac{2 c^{2} \omega^{2}}{\pi^{2}}
        f'(\varepsilon_{n_{1}}^{(0)})
        f'(\varepsilon_{n_{2}}^{(0)})
        \sin\left( \frac{A_0 - A_1}{2eB} + \varphi \right)_{n_{1}}
        \sin\left( \frac{A_0 - A_1}{2eB} + \varphi \right)_{n_{2}}
    \right]
    \\
    \bar{f}_{n} = &
    \sum_{k} f(\varepsilon_{n,k}) =
    f(\varepsilon_{n}^{(0)}) +
    f'(\varepsilon_{n}^{(0)})
    \frac{c^2 \omega}{\pi}
    (1-\alpha)
    \sin \left( \frac{A_1}{eB} - 2\varphi \right)_{n}
    +
    f''(\varepsilon_{n}^{(0)})
    \frac{c^{2} \omega^{2}}{\pi^{2}}
    \sin^{2} \left( \frac{A_0-A_1}{2eB} + \varphi \right)_{n}
\end{split}
\end{align}
\end{widetext}
Thus, $\delta \Omega$
reduces to a sum of two parts: one due to $V^{(0)}_{n_{1}, n_{2}}$ and
another due to $V^{(1)}_{n_{1}, n_{2}}$. Since
$V_{n_1, n_2}^{(0)}$ is independent of $n_1$ and $n_2$ (see Eq.
\ref{int5}), the summations over $n_{1}$ and $n_{2}$ in the
corresponding term of Eq. \ref{fact} are independent. From $V^{(0)}$ we
obtain the following contribution to $\delta \Omega$.

\begin{align}\label{Ps1}
\begin{split}
    \delta \Omega & = - \frac{e B}{\pi}
    V^{(0)}
    \left( \sum_{n} \bar{f}_{n} \right)^{2}
    \\
    \sum_{n} \bar{f}_{n} & =
    \frac{\mu}{\omega} -
    \frac{t_0 / \pi}{\text{sh}(t_0)}
    \sin \left( \frac{A_0}{eB} \right) -
    \frac{c^{2} t_1 / \pi}{\text{sh}(t_1)}
    \sin \left( \frac{A_1}{eB} - 2\varphi \right)
\end{split}
\end{align}
The second line is obtained using the Poisson summation rule and $A_0$
and $A_1$ are taken at the chemical potential. Note that the parameters
$t_0$ and $t_1$ are defined in Eq. (\ref{Oid}). Thus, upon squaring
$\sum_{n} \bar{f}_{n}$ we obtain combinations of the $A_{0}$, $A_{1}$,
and zero frequencies. This gives the interaction correction to the
standard Onsager frequencies, $A_{0}$ and $A_{1}$, present in the ideal
gas. More importantly, however, it leads to the non-Onsager frequency
$A_{0} - A_{1}$.
\begin{align}\label{nO0}
    \delta \Omega^{(0)}
    & \to
    -\frac{\lambda c^2m^2\omega^2}{4\pi^4}
    \frac{t_0}{\text{sh}(t_0)}
    \frac{t_1}{\text{sh}(t_1)} 
    \cos \left( \frac{A_0-A_1}{eB} + 2\varphi \right)
\end{align}
The non-Onsager frequency appears because the interaction contribution
is a non-linear functional of the single-particle density of states (see
Eq. \ref{int7}). This non-linearity gives rise to combinations of the
frequencies which ordinarily appear in the density of states. Note,
however, that the frequency $A_{0} - A_{1}$ should disappear at the same
temperature as $A_{0}$ according to the above equation. The term given
in Eqn. \ref{nO0} cannot explain the observed non-Onsager frequencies
which survive to `high' temperatures. To resolve this issue we turn to
the $V^{(1)}$ term in Eq. \ref{fact}.

%
%

Next, using Poisson formula, we perform the $n_1$ and $n_2$ summations
in the $V^{(1)}_{n_1n_2}$-term in (\ref{fact}). Altogether this results
in the following correction to the grand potential.
\begin{align}\label{nO1}
\begin{split}
    \delta \Omega^{(1)}
    & = - \frac{\lambda c^{2} m^{2} \omega^{2}}{2 \pi^4}
    [J_{0}(p_{F} g / e B)]^{2} \\
    & \times
    \left[
        \frac{(t_{0} - t_{1})/2}{\text{sh}((t_{0} - t_{1})/2)}
        \sin \left( \frac{A_0-A_1}{2eB} + \varphi \right)
    \right]^{2} \\
\end{split}
\end{align}
We have evaluated $M_{n}$ at $n \approx \mu / \omega \gg 1$ in which
case $M_{n} \approx J_{0}(p_{F} g / e B)$ (for Fermi momentum $p_{F}
= \sqrt{2 m \mu}$).
This interaction correction is another contribution to non-Onsager
quantum magnetic oscillations. At zero temperature, $t_{0}, t_{1} \to
0$, the contribution from $V^{(1)}$ (Eq. \ref{nO1}) is suppressed
relative to the contribution from $V^{(0)}$ (Eq. \ref{nO0}) by the
factor $[J_{0}(p_{F} g / e B)]^{2}$. Under the experimental conditions of
Ref. \cite{wang_formation_2023} this is an order of magnitude
suppression. Importantly, however, the two contributions (Eqs.
\ref{nO1} and \ref{nO0}) have a very different temperature dependence.
The non-Onsager oscillations in Eq. \ref{nO1} 
decay with temperature as $ e^{-(t_0-t_1)}=e^{-2\pi^2(1-\alpha)T/\omega}$.
At the same time Onsager oscillations decay as $A_0: e^{-2\pi^2T/\omega}$,
$A_1: e^{-2\pi^2\alpha T/\omega}$.
Typically $\alpha \approx 1$ and this explains why the non-Onsager frequency
survives till higher temperatures compared to  standard Onsager oscillations.
For conditions of the experiment \cite{wang_formation_2023} 
at electron density $\SI{1e11}{\per \square \centi \meter}$ the amplitude of the
$A_0-A_1$ oscillations decays with temperature 4 times (in exponent) slower than
the amplitude of the $A_0$ oscillation, at density   $10\times 10^{11}cm^{-2}$
the ratio is even larger, 6.


To summarise results on the present section.
We have  computed the leading-order interaction correction to the
thermodynamic potential $\delta \Omega$.
 Our major conclusions are the following. 
 (i) Electron-electron Coulomb interactions explain magnetic oscillations
with the non-Onsager frequency $A_{0} - A_{1}$. This works because
$\delta \Omega$ is a non-linear functional of the density of states.
(ii) The amplitude of the $A_{0} - A_{1}$ frequency decays with
temperature much more slowly than that of the standard Onsager
frequencies $A_{0}$ and $A_{1}$.
The interaction mechanism explains $A_0-A_1$ oscillations observed
in the de Haas–Van Alphen effect  \cite{eddy_haas---van_1982,audouard_quantum_2012}.
Of course, the mechanism also contributes to the  Shubnikov–de Haas effect.


\section{The role of chemical potential oscillations}\label{sec:muOscillations}

In this section we return to the non-interacting ideal gas.
It is well known that in gated 2D systems the chemical potential is not
constant as magnetic field is varied; the electron density is constant
instead. Technically, this means that the free energy should be used
rather than the grand potential. Practically, this means that the
chemical potential oscillates as the magnetic field is varied. Chemical
potential oscillations result in non-Onsager magnetic oscillations even
without account of electron-electron interactions
\cite{audouard_quantum_2012}. So, to complete our picture of the
non-Onsager oscillations, we present here a simple explanation of how
this occurs.

The total electron density is fixed and is equal to $n=-\partial
\Omega / \partial \mu$. We can thus find an expression for $\mu$ by
differentiating Eq. \ref{Oid}. This gives
\begin{eqnarray}\label{smu}
    && \mu = \mu_0 + \delta\mu \\
    && \delta \mu = \frac{\omega}{\pi} 
    \left[
        \frac{t_0}{\text{sh}(t_0)}
        \sin \left( \frac{A_0}{eB} \right) +
        c^2 \frac{t_1}{\text{sh}(t_1)}
        \sin \left( \frac{A_1}{eB} - 2 \varphi \right)
    \right]\nonumber
\end{eqnarray}
That is, the chemical potential can be written as constant part,
$\mu_{0}$, plus an oscillating part, $\delta \mu$, whose amplitude goes
to zero at $B = 0$. Note that $\mu_{0}$ is proportional to the total
electron density and that $A_{0}$, $A_{1}$ are evaluated at $\mu_{0}$.
The free energy reads
\begin{eqnarray}\label{free}
    F =
    \Omega( \mu_{0} + \delta \mu ) =
    \Omega(\mu_{0}) + \delta \mu \ \partial_{\mu_{0}} \Omega(\mu_{0})
\end{eqnarray}
According to our equation for the non-interacting free energy
(Eqn. \ref{Oid}) we have
\begin{align}\label{dOid}
\begin{split}
    \partial_{\mu_{0}} \Omega(\mu_{0}) = \frac{m \omega}{\pi^2}
    &
    \left[
        \frac{t_0}{\text{sh}(t_0)} \sin \frac{A_0}{eB} +
    \right. \\ & \left.
        c^2 \frac{t_1}{\text{sh}(t_1)}
        \sin \left(
                \frac{A_1}{eB}-2\varphi
             \right)
    \right]
\end{split}
 \end{align}
Hence,  the $\delta \mu \ \partial_{\mu_{0}} \Omega(\mu_{0})$ term in (\ref{free})
gives the following non-Onsager oscillation of the free energy
\begin{eqnarray}\label{df}
    \delta F =
    \frac{c^2m\omega^2}{\pi^3}\frac{t_0}{\text{sh}(t_0)}
    \frac{t_1}{\text{sh}(t_1)}
    \cos \left( \frac{A_0-A_1}{eB} + 2\varphi \right)
\end{eqnarray}
This frequency arises combinatorially; that is, because, $\Omega$
oscillates with a frequency $A_{1}$ and $\mu$ oscillates with a
frequency $A_{0}$. Note that the amplitude is even larger than that of
the interaction correction (Eq. \ref{nO0}). However, neither Eq.
\ref{df}, nor Eq. \ref{nO0} can explain the temperature dependence of
experimental data. The corresponding amplitude decays with temperature
almost twice faster than the amplitude of the main harmonic, $A_0$.
Experimentally, the amplitude of the $A_0 - A_1$ harmonic decays much
more slowly than that of $A_0$. Only the interaction effect described by
Eq. \ref{nO1} can explain this temperature dependence.

\section{Half-frequencies arising from impurities}

Magnetic oscillations with half-frequencies have been observed for the
first time in Ref. \cite{wang_formation_2023}. It was conjectured in
\cite{wang_formation_2023} that the oscillations have the kinetic
origin. Results of our Section III confirm that the conjecture was
correct,  for the triangular lattice  half-frequencies arise in
transport via the kinetic mechanism. In the present section we
demonstrate that the half non-Onsager frequency can arise even in
equilibrium in the de Haas–Van Alphen effect. In this case it is
unrelated to the electron-electron interaction, but it is related to
impurities.


Our equation for $\varepsilon_{n}^{(0)} + \delta \varepsilon_{n,k}$
(Eq. \ref{eqn:pippardSolution}) gives the energy of a quantum state in
a perfect system. While the energy correction contains a term linear in
$c$, none of the magnetic oscillations we have discussed have an
amplitude which is linear in $c$. This is because all linear terms
average to zero when we sum over the real quasi-momentum, $k$.

In the presence of impurities, however, there are also bound states and
the quasi-momentum for these states will be imaginary. We obtain the
energy of the bound states from Eq. \ref{eqn:pippardSolution} by making
the substitutions $k \rightarrow i \kappa$ and $\cos(k) \rightarrow
\text{ch}(\kappa)$, where $\kappa$ is real.
\begin{align}\label{imp1}
\begin{split}
    & \varepsilon_{n,\kappa} =
    \varepsilon_{n} ^{(0)} +
    \delta \varepsilon_{n,\kappa} \\
    & \delta \varepsilon_{n,\kappa} =
    c \frac{2 \omega}{\pi} \text{ch}(\kappa)
    \sin \left( \frac{A_0-A_1}{2eB} + \varphi \right)
\end{split}
\end{align}
To describe the half-frequencies it is sufficient to work to
leading-order in $c$. Note that according to Eq. \ref{imp1} the energy
of the localised impurity state can be above or below the mini-band of a
given Landau level, $\varepsilon_{n,k}$, depending on the sign of
$\sin( (A_{0} - A_{1}) / 2 e B + \varphi)$. Evidently, it is the
presence of this oscillating factor which leads to half-frequencies.

Averaging Eqn. \ref{imp1} over impurity states we make the replacement
$\text{ch}(\kappa) \to \langle \text{ch}(\kappa)\rangle$, where the
average $\langle \text{ch}(\kappa) \rangle$ depends on the concentration
of impurities and on the strength of the binding. The most important
point is that averages over the mini-band states, $\int \cos(k) dk$, are
zero while averages over the impurity states, $\langle \text{ch}(\kappa) \rangle$, are non-zero.

The non-zero average, $\langle \text{ch}(\kappa) \rangle$, means that
terms linear in $c$ will appear in both the single-particle density of
states and the thermodynamic potential. We focus on the thermodynamic
potential, $\Omega$, in particular. Applying the methods of section
\ref{sec:singleParticleOmega} we find that the correction to $\Omega$
due to Eqn. \ref{imp1} is

\begin{align}\label{imp3}
\begin{split}
    \delta\Omega^{(imp)}=
    \frac{2 c m\omega^2\langle \text{ch}(\kappa)\rangle}
         {\pi^3(1-\alpha)}
    &
    \left[
        \frac{(t_{0} - t_{1})/2}{\text{sh}((t_{0} - t_{1})/2)}
    \right]
    \\ \times &
    \cos \left( \frac{A_0-A_1}{2eB} + \varphi \right)
\end{split}
\end{align}
Because of the factor $\langle \text{ch}(\kappa) \rangle$ these
oscillations are weak, however, they decay with temperature much more
slowly than the main $A_{0}$ line,  $\propto e^{-(t_0-t_1)/2}=e^{-\pi^2(1-\alpha)T/\omega}$.
These two features -- in addition, of
course, to the frequency of the oscillations -- match the behaviour
observed experimentally in Ref. \cite{wang_formation_2023} It is worth
noting that the mechanism of the half-frequency $(A_{0} - A_{1})/2$
discussed in this section is distantly analogous to the AAS effect
\cite{altshuler_aaronov-bohm_1981}, in which the frequency for
Aharonov-Bohm interference is halved due to disorder.

In sense, the frequency described by Eqn. \ref{imp3} is the fundamental
fractional frequency. Given the presence of $(A_{0} - A_{1})/2$ other
frequencies, such as $(A_{0} + A_{1})/2$ and $(3 A_{0} - A_{1})/2$, also
arise. The additional frequencies appear to first order in $c$ due to
the chemical potential oscillations described in the previous section;
though in that case their temperature dependence is much stronger than
$(A_{0} - A_{1})/2$ (see the discussion at the end of section
\ref{sec:muOscillations}). These additional half-frequencies were
observed in Ref. \cite{wang_formation_2023} with a stronger temperature
dependence compared to $(A_{0} - A_{1})/2$.

Since most of the oscillations effects have a geometrical interpretation
it is worth mentioning the geometrical interpretation of $(A_{0} -
A_{1})/2$. This frequency is related to the phase, $(A_{0} - A_{1})/2 e
B$, acquired along one period of the open trajectory in Fig.
\ref{fig:scat}. That the oscillations in Eqn. \ref{imp3} appear to
linear order in $c$ relates to the fact that this trajectory contains a
single reflection from magnetic breakdown junction.



\section{Conclusions}

Standard magnetic quantum oscillations are explained by Bohr-Sommerfeld
quantization of electron dynamics along closed electron trajectories in
a magnetic field. The oscillations are periodic in inverse magnetic
field and, according to Onsager, the frequency of the oscillations is
equal to the area of the closed electron trajectory. We refer to this
single-particle effect as ``Onsager oscillations''. For an electron in a
periodic potential (crystal lattice or artificial superlattice) there
are multiple closed trajectories and hence there are multiple Onsager
frequencies. Oscillations with frequencies which do not correspond to
any closed electron trajectory have also been observed and have remained
not fully understood for decades. We call these oscillations
``non-Onsager oscillations''. We show that there are two mechanisms
responsible for these oscillations: (i) single particle kinetics, and
(ii)  multi-electron correlations. Both mechanisms contribute to the
Shubnikov–de Haas effect and only the second mechanism contributes to
the de Haas–Van Alphen effect. Very recently, half-frequency non-Onsager
quantum magnetic oscillations were discovered in measurements with a two
dimensional electron gas in GaAs on a superlattice. We show that the
half-frequency oscillations also have two mechanisms: (i) single
particle kinetics, and (ii) single particle dynamics with account of
impurities. Again, both mechanisms contribute to the Shubnikov–de Haas
effect and only the second mechanism contributes to the De Haas–Van
Alphen effect.

We have developed a comprehensive theory of non-Onsager quantum magnetic
oscillations in the magnetic breakdown regime. Non-Onsager means that
the frequency of these oscillation is equal to the area of a trajectory
that an electron cannot transverse.  Hence, they are not directly
related to the Onsager quantization. Experimentally integer non-Onsager
oscillations have been known for quite some time. Very recently
half-frequency non-Onsager oscillations have also been observed.

We demonstrate that there are two mechanisms for   the integer
oscillations, (i) single particle kinetics, (ii) multi-electron
correlations due to the Coulomb electron-electron interaction. The both
mechanisms contribute to the Shubnikov–de Haas effect (transport) and
only the second mechanism  contributes to the De Haas–Van Alphen effect
(equilibrium). We calculate the temperature dependence of quantum
non-Onsager oscillations and hence explain why they survive up to
relatively high temperatures in spite of being fully quantum. 

There are also two mechanisms for the half integer oscillations, both
mechanisms are  single particle (i) single particle kinetics, (ii)
single particle dynamics with account of impurities. The both mechanisms
contribute to the Shubnikov–de Haas effect (transport) and only the
second mechanism contributes to the De Haas–Van Alphen effect
(equilibrium).

To summarise, we explain the newly discovered half-frequency non-Onsager
quantum magnetic oscillations and resolve the long standing theoretical
controversy of integer non-Onsager quantum magnetic oscillations.

\begin{acknowledgements}

This work was funded by the Australian Research Council Centre of
Excellence in Future Low-Energy Electronics Technology (FLEET) (Grant
No. CE170100039).

\end{acknowledgements}

\appendix

\section{Theory of the scattering phase in magnetic breakdown}

To compute the scattering phase we first map the scattering problem to
the Landau-Zener problem (Ref. \cite{zener_non-adiabatic_1932}) and then
make use of Zener's result \cite{zener_non-adiabatic_1932} to obtain the
scattering phase.

Semi-classical orbits exist in mechanical momentum space, defined by
$[k_{x}, k_{y}] = i e B$. This is the space in which the circular orbit
of Fig. \ref{fig:intro} lives. At the intersection between two orbits we
have two bands coupled by an off-diagonal matrix element $W$. The
Hamiltonian is

\begin{align*}
    H =
    \begin{bmatrix}
        ( (k_{x} - g )^{2} + k_{y}^{2} )/2m & W \\
        W & ( k_{x}^{2} + k_{y}^{2} )/2m
    \end{bmatrix}
\end{align*}

We then linearise this Hamiltonian around the intersection point
$\bm{k}_{0} = (g/2, \sqrt{k_{F}^{2} - g^{2} / 4})$ (see Fig.
\ref{fig:scat}a) and impose an on-shell condition, $H \psi =
\varepsilon_{F} \psi$. If we define $\bm{k} = \bm{k}_{0} + (e B X, P)$
then $X$ and $P$ give the deviation from the intersection point and are
canonically conjugate. When the condition $H \psi = \varepsilon_{F}
\psi$ is made linear in the variables $X$ and $P$ we obtain

\begin{align}\label{equ:CMBD_landauZenerEffectiveHamiltonian}
    - i \partial_{X} \psi =
    \begin{bmatrix}
        + \frac{e B k_{x0}}{k_{y0}} X & - \frac{m W}{k_{y0}} \\
        - \frac{m W}{k_{y0}} & - \frac{e B k_{x0}}{k_{y0}} X
    \end{bmatrix} \psi
    = \mathcal{H} \psi
\end{align}

Which follows from $P = -i \partial_{X}$. The above, effective
Hamiltonian, $\mathcal{H}$ has the same form as the Hamiltonian used by
Zener in Ref. \cite{zener_non-adiabatic_1932}. Re-writing Eqn.
\ref{equ:CMBD_landauZenerEffectiveHamiltonian} in Zener's notation gives

\begin{align*}
    \mathcal{H} =
    \begin{bmatrix}
        + \frac{\alpha}{2} t & f \\
        f &  - \frac{\alpha}{2} t
    \end{bmatrix}
\end{align*}

In this notation $t = X$, $\alpha > 0$ and $f < 0$. The effective
Schr\"odinger equation $\mathcal{H} \psi = - i \partial_{t} \psi$ has
solutions that we write in the form

\begin{align*}
    \psi =
    \begin{bmatrix}
        C_{1}(t) e^{+ i \alpha t^{2} / 4} \\
        C_{3}(t) e^{- i \alpha t^{2} / 4}
    \end{bmatrix}
\end{align*}

Here, the component $C_{1}$ corresponds to the wave 1 in Fig.
\ref{fig:scat} and the component $C_{3}$ corresponds to the wave 3 in
the same figure. Both waves are propagating towards the intersection
point. Suppose that the particle is initially in the state 3 of Fig.
\ref{fig:scat}a.

\begin{align*}
     C_{1}(-\infty)  &= 0 \\
    |C_{3}(-\infty)| &= 1
\end{align*}

The scattering phase is determined by the components at time $t =
\infty$: $C_{1}(\infty)$ (which relates to transitions $3 \to 4$),
$C_{3}(\infty)$ (which relates to transitions $3 \to 2$). The full
details of the calculation are contained in Ref.
\cite{zener_non-adiabatic_1932}, which describes how to reduce the
effective Schr\"odinger equation to the Weber equation
\cite{whittaker_confluent_1996}. In the end the asymptotic wavevector
amplitudes are determined by the parabolic cylinder functions (Eqns.
9.246 in Ref. \cite{zwillinger_9_2014}). The amplitudes are summarised
in table \ref{tab:CMBD_asympSummary}.

In table \ref{tab:CMBD_asympSummary} we have $n = i f^{2} / \alpha$.
Since $n$ is pure imaginary most of the factors in the these expressions
have unit magnitude. Consistent with our boundary conditions we have
$|C_{1}(-\infty)| = 0$ and $|C_{3}(-\infty)| = 1$. And consistent with
our expectations we have $|C_{1}(\infty)| = c$, and $|C_{3}(\infty)| =
s$, where $s$ is the probability of magnetic breakdown $s = \exp(-2 \pi
f^{2} / \alpha)$ (equivalent to Eqn. \ref{scpar}). In addition to
information about the probability of magnetic breakdown, table
\ref{tab:CMBD_asympSummary} also contains information about the phase
acquired during a quantum jump. For the transitions defined in Fig.
\ref{fig:scat}a we have

\begin{align*}
    3 & \rightarrow 2  \implies \Delta \phi = 0 \\
    3 & \rightarrow 4  \implies \Delta \phi = \phi_{R}
\end{align*}

Where $\phi_{R}$ is the phase acquired upon reflection from a magnetic breakdown junction and is given by

\begin{align}\label{equ:CMBD_reflectionPhase}
    \phi_{R} =
    + \frac{\pi}{4}
    - \arg \left(
               \Gamma \left( 1 + i \frac{f^{2}}{\alpha} \right)
           \right)
    + 2\frac{f^{2}}{\alpha}
       \ln \left( \sqrt{\alpha} |t| \right)
\end{align}

Here $\Gamma$ is the gamma function, and the variables $f$ and $\alpha$ are defined by

\begin{align*}
    f =
    \frac{m W}{ k_{F} \sqrt{1 - (\frac{g}{2 k_{F}})^{2}}}
    \quad \quad \quad \quad
    \alpha =
    \frac{e B g}{k_{F}\sqrt{1 - (\frac{g}{2 k_{F}})^{2}} }
\end{align*}

The phase $\phi_{R}$ contains terms which are of the form $A / e B$;
that is, terms which simply give a correction to the areas of the Fermi
surface. The true scattering phase should not contain these area
corrections and should equal zero in the adiabatic limit (large $W$ or
$f^{2} / \alpha \to \infty$). The expression satisfying these conditions
is

\begin{align}\label{equ:CMBD_trueScatteringPhase}
    \varphi & =
    + \frac{\pi}{4}
    - \arg \left(
               \Gamma \left( 1 + i \frac{f^{2}}{\alpha} \right)
           \right)
    + \frac{f^{2}}{\alpha}
      \left( \ln \left( \frac{f^{2}}{\alpha} \right) - 1 \right)
\end{align}

\begin{table*}[hb]
\center
\begin{tabular}{|| c | c | c ||}
    \hline
    $C$ & $t \rightarrow - \infty$ & $t \rightarrow + \infty$ \\
    [0.5ex] \hline\hline
    $C_{1}$ & \quad
    $\alpha^{-\frac{n}{2}}
    |t|^{-n}
    e^{- i \frac{\pi}{4}}
    e^{- i \alpha t^{2} / 2}
    \frac{f}{\alpha |t|}$ \quad & \quad
    $e^{- i \arg(\Gamma(n+1)) }
    \alpha^{\frac{n}{2}}
    |t|^{n}
    (1 - P)^{1/2}$ \quad \\
    [0.5ex] \hline
    $C_{3}$ & \quad
    $\alpha^{-\frac{n}{2}}
    |t|^{-n}
    e^{- i \frac{\pi}{4}}$ \quad & \quad
    $\alpha^{-\frac{n}{2}}
    |t|^{-n}
    e^{- i \frac{\pi}{4}}
    P^{1/2}$ \quad \\
    [0.5ex] \hline
\end{tabular}
\caption{Summary of the asymptotic components of the wavevector $\psi$.}
\label{tab:CMBD_asympSummary}
\end{table*}


\begin{thebibliography}{34}%
\makeatletter
\providecommand \@ifxundefined [1]{%
 \@ifx{#1\undefined}
}%
\providecommand \@ifnum [1]{%
 \ifnum #1\expandafter \@firstoftwo
 \else \expandafter \@secondoftwo
 \fi
}%
\providecommand \@ifx [1]{%
 \ifx #1\expandafter \@firstoftwo
 \else \expandafter \@secondoftwo
 \fi
}%
\providecommand \natexlab [1]{#1}%
\providecommand \enquote  [1]{``#1''}%
\providecommand \bibnamefont  [1]{#1}%
\providecommand \bibfnamefont [1]{#1}%
\providecommand \citenamefont [1]{#1}%
\providecommand \href@noop [0]{\@secondoftwo}%
\providecommand \href [0]{\begingroup \@sanitize@url \@href}%
\providecommand \@href[1]{\@@startlink{#1}\@@href}%
\providecommand \@@href[1]{\endgroup#1\@@endlink}%
\providecommand \@sanitize@url [0]{\catcode `\\12\catcode `\$12\catcode `\&12\catcode `\#12\catcode `\^12\catcode `\_12\catcode `\%12\relax}%
\providecommand \@@startlink[1]{}%
\providecommand \@@endlink[0]{}%
\providecommand \url  [0]{\begingroup\@sanitize@url \@url }%
\providecommand \@url [1]{\endgroup\@href {#1}{\urlprefix }}%
\providecommand \urlprefix  [0]{URL }%
\providecommand \Eprint [0]{\href }%
\providecommand \doibase [0]{https://doi.org/}%
\providecommand \selectlanguage [0]{\@gobble}%
\providecommand \bibinfo  [0]{\@secondoftwo}%
\providecommand \bibfield  [0]{\@secondoftwo}%
\providecommand \translation [1]{[#1]}%
\providecommand \BibitemOpen [0]{}%
\providecommand \bibitemStop [0]{}%
\providecommand \bibitemNoStop [0]{.\EOS\space}%
\providecommand \EOS [0]{\spacefactor3000\relax}%
\providecommand \BibitemShut  [1]{\csname bibitem#1\endcsname}%
\let\auto@bib@innerbib\@empty
\bibitem [{\citenamefont {Abrikosov}(2017)}]{abrikosov_fundamentals_2017}%
  \BibitemOpen
  \bibfield  {author} {\bibinfo {author} {\bibfnamefont {A.}~\bibnamefont {Abrikosov}},\ }\href@noop {} {\emph {\bibinfo {title} {Fundamentals of the {Theory} of {Metals}}}}\ (\bibinfo  {publisher} {Dover Publications},\ \bibinfo {year} {2017})\BibitemShut {NoStop}%
\bibitem [{\citenamefont {Priestley}\ \emph {et~al.}(1963)\citenamefont {Priestley}, \citenamefont {Falicov},\ and\ \citenamefont {Weisz}}]{priestley_experimental_1963}%
  \BibitemOpen
  \bibfield  {author} {\bibinfo {author} {\bibfnamefont {M.~G.}\ \bibnamefont {Priestley}}, \bibinfo {author} {\bibfnamefont {L.~M.}\ \bibnamefont {Falicov}},\ and\ \bibinfo {author} {\bibfnamefont {G.}~\bibnamefont {Weisz}},\ }\bibfield  {title} {\bibinfo {title} {Experimental and {Theoretical} {Study} of {Magnetic} {Breakdown} in {Magnesium}},\ }\href {https://doi.org/10.1103/PhysRev.131.617} {\bibfield  {journal} {\bibinfo  {journal} {Physical Review}\ }\textbf {\bibinfo {volume} {131}},\ \bibinfo {pages} {617} (\bibinfo {year} {1963})},\ \bibinfo {note} {publisher: American Physical Society}\BibitemShut {NoStop}%
\bibitem [{\citenamefont {Alexandradinata}\ and\ \citenamefont {Glazman}(2017)}]{alexandradinata_geometric_2017}%
  \BibitemOpen
  \bibfield  {author} {\bibinfo {author} {\bibfnamefont {A.}~\bibnamefont {Alexandradinata}}\ and\ \bibinfo {author} {\bibfnamefont {L.}~\bibnamefont {Glazman}},\ }\bibfield  {title} {\bibinfo {title} {Geometric {Phase} and {Orbital} {Moment} in {Quantization} {Rules} for {Magnetic} {Breakdown}},\ }\href {https://doi.org/10.1103/PhysRevLett.119.256601} {\bibfield  {journal} {\bibinfo  {journal} {Physical Review Letters}\ }\textbf {\bibinfo {volume} {119}},\ \bibinfo {pages} {256601} (\bibinfo {year} {2017})},\ \bibinfo {note} {publisher: American Physical Society}\BibitemShut {NoStop}%
\bibitem [{\citenamefont {Alexandradinata}\ and\ \citenamefont {Glazman}(2023)}]{alexandradinata_fermiology_2023}%
  \BibitemOpen
  \bibfield  {author} {\bibinfo {author} {\bibfnamefont {A.}~\bibnamefont {Alexandradinata}}\ and\ \bibinfo {author} {\bibfnamefont {L.}~\bibnamefont {Glazman}},\ }\bibfield  {title} {\bibinfo {title} {Fermiology of {Topological} {Metals}},\ }\href {https://doi.org/10.1146/annurev-conmatphys-040721-021331} {\bibfield  {journal} {\bibinfo  {journal} {Annual Review of Condensed Matter Physics}\ }\textbf {\bibinfo {volume} {14}},\ \bibinfo {pages} {261} (\bibinfo {year} {2023})},\ \bibinfo {note} {\_eprint: https://doi.org/10.1146/annurev-conmatphys-040721-021331}\BibitemShut {NoStop}%
\bibitem [{\citenamefont {Kunisada}\ \emph {et~al.}(2020)\citenamefont {Kunisada}, \citenamefont {Isono}, \citenamefont {Kohama}, \citenamefont {Sakai}, \citenamefont {Bareille}, \citenamefont {Sakuragi}, \citenamefont {Noguchi}, \citenamefont {Kurokawa}, \citenamefont {Kuroda}, \citenamefont {Ishida}, \citenamefont {Adachi}, \citenamefont {Sekine}, \citenamefont {Kim}, \citenamefont {Cacho}, \citenamefont {Shin}, \citenamefont {Tohyama}, \citenamefont {Tokiwa},\ and\ \citenamefont {Kondo}}]{kunisada2020}%
  \BibitemOpen
  \bibfield  {author} {\bibinfo {author} {\bibfnamefont {S.}~\bibnamefont {Kunisada}}, \bibinfo {author} {\bibfnamefont {S.}~\bibnamefont {Isono}}, \bibinfo {author} {\bibfnamefont {Y.}~\bibnamefont {Kohama}}, \bibinfo {author} {\bibfnamefont {S.}~\bibnamefont {Sakai}}, \bibinfo {author} {\bibfnamefont {C.}~\bibnamefont {Bareille}}, \bibinfo {author} {\bibfnamefont {S.}~\bibnamefont {Sakuragi}}, \bibinfo {author} {\bibfnamefont {R.}~\bibnamefont {Noguchi}}, \bibinfo {author} {\bibfnamefont {K.}~\bibnamefont {Kurokawa}}, \bibinfo {author} {\bibfnamefont {K.}~\bibnamefont {Kuroda}}, \bibinfo {author} {\bibfnamefont {Y.}~\bibnamefont {Ishida}}, \bibinfo {author} {\bibfnamefont {S.}~\bibnamefont {Adachi}}, \bibinfo {author} {\bibfnamefont {R.}~\bibnamefont {Sekine}}, \bibinfo {author} {\bibfnamefont {T.~K.}\ \bibnamefont {Kim}}, \bibinfo {author} {\bibfnamefont {C.}~\bibnamefont {Cacho}}, \bibinfo {author} {\bibfnamefont {S.}~\bibnamefont {Shin}}, \bibinfo {author} {\bibfnamefont {T.}~\bibnamefont {Tohyama}}, \bibinfo {author} {\bibfnamefont {K.}~\bibnamefont {Tokiwa}},\ and\ \bibinfo {author} {\bibfnamefont {T.}~\bibnamefont {Kondo}},\ }\bibfield  {title} {\bibinfo {title} {Observation of small fermi pockets protected by clean {CuO2} sheets of a high-{Tc} superconductor},\ }\href {https://doi.org/10.1126/science.aay7311} {\bibfield  {journal} {\bibinfo  {journal} {Science}\ }\textbf {\bibinfo {volume} {369}},\ \bibinfo {pages} {833} (\bibinfo {year} {2020})},\ \Eprint {https://arxiv.org/abs/https://www.science.org/doi/pdf/10.1126/science.aay7311} {https://www.science.org/doi/pdf/10.1126/science.aay7311} \BibitemShut {NoStop}%
\bibitem [{\citenamefont {Kurokawa}\ \emph {et~al.}(2023)\citenamefont {Kurokawa}, \citenamefont {Isono}, \citenamefont {Kohama}, \citenamefont {Kunisada}, \citenamefont {Sakai}, \citenamefont {Sekine}, \citenamefont {Okubo}, \citenamefont {Watson}, \citenamefont {Kim}, \citenamefont {Cacho}, \citenamefont {Shin}, \citenamefont {Tohyama}, \citenamefont {Tokiwa},\ and\ \citenamefont {Kondo}}]{kurokawa_unveiling_2023}%
  \BibitemOpen
  \bibfield  {author} {\bibinfo {author} {\bibfnamefont {K.}~\bibnamefont {Kurokawa}}, \bibinfo {author} {\bibfnamefont {S.}~\bibnamefont {Isono}}, \bibinfo {author} {\bibfnamefont {Y.}~\bibnamefont {Kohama}}, \bibinfo {author} {\bibfnamefont {S.}~\bibnamefont {Kunisada}}, \bibinfo {author} {\bibfnamefont {S.}~\bibnamefont {Sakai}}, \bibinfo {author} {\bibfnamefont {R.}~\bibnamefont {Sekine}}, \bibinfo {author} {\bibfnamefont {M.}~\bibnamefont {Okubo}}, \bibinfo {author} {\bibfnamefont {M.~D.}\ \bibnamefont {Watson}}, \bibinfo {author} {\bibfnamefont {T.~K.}\ \bibnamefont {Kim}}, \bibinfo {author} {\bibfnamefont {C.}~\bibnamefont {Cacho}}, \bibinfo {author} {\bibfnamefont {S.}~\bibnamefont {Shin}}, \bibinfo {author} {\bibfnamefont {T.}~\bibnamefont {Tohyama}}, \bibinfo {author} {\bibfnamefont {K.}~\bibnamefont {Tokiwa}},\ and\ \bibinfo {author} {\bibfnamefont {T.}~\bibnamefont {Kondo}},\ }\bibfield  {title} {{\selectlanguage {english}\bibinfo {title} {Unveiling phase diagram of the lightly doped high-{Tc} cuprate superconductors with disorder removed}},\ }\href {https://doi.org/10.1038/s41467-023-39457-7} {\bibfield  {journal} {\bibinfo  {journal} {Nature Communications}\ }\textbf {\bibinfo {volume} {14}},\ \bibinfo {pages} {4064} (\bibinfo {year} {2023})},\ \bibinfo {note} {number: 1 Publisher: Nature Publishing Group}\BibitemShut {NoStop}%
\bibitem [{\citenamefont {de~Vries}\ \emph {et~al.}(2023)\citenamefont {de~Vries}, \citenamefont {Slizovskiy}, \citenamefont {Tomić}, \citenamefont {Kumar}, \citenamefont {Garcia-Ruiz}, \citenamefont {Zheng}, \citenamefont {Portolés}, \citenamefont {Ponomarenko}, \citenamefont {Geim}, \citenamefont {Watanabe}, \citenamefont {Taniguchi}, \citenamefont {Fal'ko}, \citenamefont {Ensslin}, \citenamefont {Ihn},\ and\ \citenamefont {Rickhaus}}]{de_vries_kagome_2023}%
  \BibitemOpen
  \bibfield  {author} {\bibinfo {author} {\bibfnamefont {F.~K.}\ \bibnamefont {de~Vries}}, \bibinfo {author} {\bibfnamefont {S.}~\bibnamefont {Slizovskiy}}, \bibinfo {author} {\bibfnamefont {P.}~\bibnamefont {Tomić}}, \bibinfo {author} {\bibfnamefont {R.~K.}\ \bibnamefont {Kumar}}, \bibinfo {author} {\bibfnamefont {A.}~\bibnamefont {Garcia-Ruiz}}, \bibinfo {author} {\bibfnamefont {G.}~\bibnamefont {Zheng}}, \bibinfo {author} {\bibfnamefont {E.}~\bibnamefont {Portolés}}, \bibinfo {author} {\bibfnamefont {L.~A.}\ \bibnamefont {Ponomarenko}}, \bibinfo {author} {\bibfnamefont {A.~K.}\ \bibnamefont {Geim}}, \bibinfo {author} {\bibfnamefont {K.}~\bibnamefont {Watanabe}}, \bibinfo {author} {\bibfnamefont {T.}~\bibnamefont {Taniguchi}}, \bibinfo {author} {\bibfnamefont {V.}~\bibnamefont {Fal'ko}}, \bibinfo {author} {\bibfnamefont {K.}~\bibnamefont {Ensslin}}, \bibinfo {author} {\bibfnamefont {T.}~\bibnamefont {Ihn}},\ and\ \bibinfo {author} {\bibfnamefont {P.}~\bibnamefont {Rickhaus}},\ }\href {https://arxiv.org/abs/2303.06403v2} {{\selectlanguage {english}\bibinfo {title} {Kagome quantum oscillations in graphene superlattices}}} (\bibinfo {year} {2023})\BibitemShut {NoStop}%
\bibitem [{\citenamefont {Wang}\ \emph {et~al.}(2023)\citenamefont {Wang}, \citenamefont {Krix}, \citenamefont {Sushkov}, \citenamefont {Farrer}, \citenamefont {Ritchie}, \citenamefont {Hamilton},\ and\ \citenamefont {Klochan}}]{wang_formation_2023}%
  \BibitemOpen
  \bibfield  {author} {\bibinfo {author} {\bibfnamefont {D.~Q.}\ \bibnamefont {Wang}}, \bibinfo {author} {\bibfnamefont {Z.}~\bibnamefont {Krix}}, \bibinfo {author} {\bibfnamefont {O.~P.}\ \bibnamefont {Sushkov}}, \bibinfo {author} {\bibfnamefont {I.}~\bibnamefont {Farrer}}, \bibinfo {author} {\bibfnamefont {D.~A.}\ \bibnamefont {Ritchie}}, \bibinfo {author} {\bibfnamefont {A.~R.}\ \bibnamefont {Hamilton}},\ and\ \bibinfo {author} {\bibfnamefont {O.}~\bibnamefont {Klochan}},\ }\bibfield  {title} {\bibinfo {title} {Formation of {Artificial} {Fermi} {Surfaces} with a {Triangular} {Superlattice} on a {Conventional} {Two}-{Dimensional} {Electron} {Gas}},\ }\bibfield  {journal} {\bibinfo  {journal} {Nano Letters}\ }\href {https://doi.org/10.1021/acs.nanolett.2c04358} {10.1021/acs.nanolett.2c04358} (\bibinfo {year} {2023}),\ \bibinfo {note} {publisher: American Chemical Society}\BibitemShut {NoStop}%
\bibitem [{\citenamefont {Onsager}(1952)}]{onsager_interpretation_1952}%
  \BibitemOpen
  \bibfield  {author} {\bibinfo {author} {\bibfnamefont {L.}~\bibnamefont {Onsager}},\ }\bibfield  {title} {\bibinfo {title} {Interpretation of the de {Haas}-van {Alphen} effect},\ }\href {https://doi.org/10.1080/14786440908521019} {\bibfield  {journal} {\bibinfo  {journal} {The London, Edinburgh, and Dublin Philosophical Magazine and Journal of Science}\ }\textbf {\bibinfo {volume} {43}},\ \bibinfo {pages} {1006} (\bibinfo {year} {1952})}\BibitemShut {NoStop}%
\bibitem [{\citenamefont {Wang}\ \emph {et~al.}(2020)\citenamefont {Wang}, \citenamefont {Reuter}, \citenamefont {Wieck}, \citenamefont {Hamilton},\ and\ \citenamefont {Klochan}}]{wang_two-dimensional_2020}%
  \BibitemOpen
  \bibfield  {author} {\bibinfo {author} {\bibfnamefont {D.~Q.}\ \bibnamefont {Wang}}, \bibinfo {author} {\bibfnamefont {D.}~\bibnamefont {Reuter}}, \bibinfo {author} {\bibfnamefont {A.~D.}\ \bibnamefont {Wieck}}, \bibinfo {author} {\bibfnamefont {A.~R.}\ \bibnamefont {Hamilton}},\ and\ \bibinfo {author} {\bibfnamefont {O.}~\bibnamefont {Klochan}},\ }\bibfield  {title} {\bibinfo {title} {Two-dimensional lateral surface superlattices in {GaAs} heterostructures with independent control of carrier density and modulation potential},\ }\href {https://doi.org/10.1063/5.0009462} {\bibfield  {journal} {\bibinfo  {journal} {Applied Physics Letters}\ }\textbf {\bibinfo {volume} {117}},\ \bibinfo {pages} {032102} (\bibinfo {year} {2020})}\BibitemShut {NoStop}%
\bibitem [{\citenamefont {Stark}\ and\ \citenamefont {Friedberg}(1974)}]{stark_interfering_1974}%
  \BibitemOpen
  \bibfield  {author} {\bibinfo {author} {\bibfnamefont {R.~W.}\ \bibnamefont {Stark}}\ and\ \bibinfo {author} {\bibfnamefont {C.~B.}\ \bibnamefont {Friedberg}},\ }\bibfield  {title} {{\selectlanguage {english}\bibinfo {title} {Interfering electron quantum states in ultrapure magnesium}},\ }\href {https://doi.org/10.1007/BF00654814} {\bibfield  {journal} {\bibinfo  {journal} {Journal of Low Temperature Physics}\ }\textbf {\bibinfo {volume} {14}},\ \bibinfo {pages} {111} (\bibinfo {year} {1974})}\BibitemShut {NoStop}%
\bibitem [{\citenamefont {Morrison}\ and\ \citenamefont {Stark}(1981)}]{morrison_two-lifetime_1981}%
  \BibitemOpen
  \bibfield  {author} {\bibinfo {author} {\bibfnamefont {D.}~\bibnamefont {Morrison}}\ and\ \bibinfo {author} {\bibfnamefont {R.~W.}\ \bibnamefont {Stark}},\ }\bibfield  {title} {{\selectlanguage {english}\bibinfo {title} {Two-lifetime model calculations of the quantum interference dominated transverse magnetoresistance of magnesium}},\ }\href {https://doi.org/10.1007/BF00654499} {\bibfield  {journal} {\bibinfo  {journal} {Journal of Low Temperature Physics}\ }\textbf {\bibinfo {volume} {45}},\ \bibinfo {pages} {531} (\bibinfo {year} {1981})}\BibitemShut {NoStop}%
\bibitem [{\citenamefont {Gerhardts}\ \emph {et~al.}(1989)\citenamefont {Gerhardts}, \citenamefont {Weiss},\ and\ \citenamefont {Klitzing}}]{gerhardts_novel_1989}%
  \BibitemOpen
  \bibfield  {author} {\bibinfo {author} {\bibfnamefont {R.~R.}\ \bibnamefont {Gerhardts}}, \bibinfo {author} {\bibfnamefont {D.}~\bibnamefont {Weiss}},\ and\ \bibinfo {author} {\bibfnamefont {K.~v.}\ \bibnamefont {Klitzing}},\ }\bibfield  {title} {\bibinfo {title} {Novel magnetoresistance oscillations in a periodically modulated two-dimensional electron gas},\ }\href {https://doi.org/10.1103/PhysRevLett.62.1173} {\bibfield  {journal} {\bibinfo  {journal} {Physical Review Letters}\ }\textbf {\bibinfo {volume} {62}},\ \bibinfo {pages} {1173} (\bibinfo {year} {1989})}\BibitemShut {NoStop}%
\bibitem [{\citenamefont {Deutschmann}\ \emph {et~al.}(2001)\citenamefont {Deutschmann}, \citenamefont {Wegscheider}, \citenamefont {Rother}, \citenamefont {Bichler}, \citenamefont {Abstreiter}, \citenamefont {Albrecht},\ and\ \citenamefont {Smet}}]{deutschmann_quantum_2001}%
  \BibitemOpen
  \bibfield  {author} {\bibinfo {author} {\bibfnamefont {R.~A.}\ \bibnamefont {Deutschmann}}, \bibinfo {author} {\bibfnamefont {W.}~\bibnamefont {Wegscheider}}, \bibinfo {author} {\bibfnamefont {M.}~\bibnamefont {Rother}}, \bibinfo {author} {\bibfnamefont {M.}~\bibnamefont {Bichler}}, \bibinfo {author} {\bibfnamefont {G.}~\bibnamefont {Abstreiter}}, \bibinfo {author} {\bibfnamefont {C.}~\bibnamefont {Albrecht}},\ and\ \bibinfo {author} {\bibfnamefont {J.~H.}\ \bibnamefont {Smet}},\ }\bibfield  {title} {\bibinfo {title} {Quantum {Interference} in {Artificial} {Band} {Structures}},\ }\href {https://doi.org/10.1103/PhysRevLett.86.1857} {\bibfield  {journal} {\bibinfo  {journal} {Physical Review Letters}\ }\textbf {\bibinfo {volume} {86}},\ \bibinfo {pages} {1857} (\bibinfo {year} {2001})},\ \bibinfo {note} {publisher: American Physical Society}\BibitemShut {NoStop}%
\bibitem [{\citenamefont {Meyer}\ \emph {et~al.}(1995)\citenamefont {Meyer}, \citenamefont {Steep}, \citenamefont {Biberacher}, \citenamefont {Christ}, \citenamefont {Lerf}, \citenamefont {Jansen}, \citenamefont {Joss}, \citenamefont {Wyder},\ and\ \citenamefont {Andres}}]{meyer_high-field_1995}%
  \BibitemOpen
  \bibfield  {author} {\bibinfo {author} {\bibfnamefont {F.~A.}\ \bibnamefont {Meyer}}, \bibinfo {author} {\bibfnamefont {E.}~\bibnamefont {Steep}}, \bibinfo {author} {\bibfnamefont {W.}~\bibnamefont {Biberacher}}, \bibinfo {author} {\bibfnamefont {P.}~\bibnamefont {Christ}}, \bibinfo {author} {\bibfnamefont {A.}~\bibnamefont {Lerf}}, \bibinfo {author} {\bibfnamefont {A.~G.~M.}\ \bibnamefont {Jansen}}, \bibinfo {author} {\bibfnamefont {W.}~\bibnamefont {Joss}}, \bibinfo {author} {\bibfnamefont {P.}~\bibnamefont {Wyder}},\ and\ \bibinfo {author} {\bibfnamefont {K.}~\bibnamefont {Andres}},\ }\bibfield  {title} {{\selectlanguage {english}\bibinfo {title} {High-{Field} de {Haas}-{Van} {Alphen} {Studies} of kappa-({BEDT}-{TTF}){2Cu}({NCS})2}},\ }\href {https://doi.org/10.1209/0295-5075/32/8/011} {\bibfield  {journal} {\bibinfo  {journal} {Europhysics Letters}\ }\textbf {\bibinfo {volume} {32}},\ \bibinfo {pages} {681} (\bibinfo {year} {1995})}\BibitemShut {NoStop}%
\bibitem [{\citenamefont {Audouard}\ \emph {et~al.}(2012)\citenamefont {Audouard}, \citenamefont {Fortin}, \citenamefont {Vignolles}, \citenamefont {Lyubovskii}, \citenamefont {Drigo}, \citenamefont {Duc}, \citenamefont {Shilov}, \citenamefont {Ballon}, \citenamefont {Zhilyaeva}, \citenamefont {Lyubovskaya},\ and\ \citenamefont {Canadell}}]{audouard_quantum_2012}%
  \BibitemOpen
  \bibfield  {author} {\bibinfo {author} {\bibfnamefont {A.}~\bibnamefont {Audouard}}, \bibinfo {author} {\bibfnamefont {J.-Y.}\ \bibnamefont {Fortin}}, \bibinfo {author} {\bibfnamefont {D.}~\bibnamefont {Vignolles}}, \bibinfo {author} {\bibfnamefont {R.~B.}\ \bibnamefont {Lyubovskii}}, \bibinfo {author} {\bibfnamefont {L.}~\bibnamefont {Drigo}}, \bibinfo {author} {\bibfnamefont {F.}~\bibnamefont {Duc}}, \bibinfo {author} {\bibfnamefont {G.~V.}\ \bibnamefont {Shilov}}, \bibinfo {author} {\bibfnamefont {G.}~\bibnamefont {Ballon}}, \bibinfo {author} {\bibfnamefont {E.~I.}\ \bibnamefont {Zhilyaeva}}, \bibinfo {author} {\bibfnamefont {R.~N.}\ \bibnamefont {Lyubovskaya}},\ and\ \bibinfo {author} {\bibfnamefont {E.}~\bibnamefont {Canadell}},\ }\bibfield  {title} {{\selectlanguage {english}\bibinfo {title} {Quantum oscillations in the linear chain of coupled orbits: {The} organic metal with two cation layers theta-({ET}){4CoBr4}({C6H4Cl2})}},\ }\href {https://doi.org/10.1209/0295-5075/97/57003} {\bibfield  {journal} {\bibinfo  {journal} {EPL (Europhysics Letters)}\ }\textbf {\bibinfo {volume} {97}},\ \bibinfo {pages} {57003} (\bibinfo {year} {2012})},\ \bibinfo {note} {publisher: IOP Publishing}\BibitemShut {NoStop}%
\bibitem [{\citenamefont {Beenakker}(1989)}]{beenakker_guiding-center-drift_1989}%
  \BibitemOpen
  \bibfield  {author} {\bibinfo {author} {\bibfnamefont {C.~W.~J.}\ \bibnamefont {Beenakker}},\ }\bibfield  {title} {\bibinfo {title} {Guiding-center-drift resonance in a periodically modulated two-dimensional electron gas},\ }\href {https://doi.org/10.1103/PhysRevLett.62.2020} {\bibfield  {journal} {\bibinfo  {journal} {Physical Review Letters}\ }\textbf {\bibinfo {volume} {62}},\ \bibinfo {pages} {2020} (\bibinfo {year} {1989})},\ \bibinfo {note} {publisher: American Physical Society}\BibitemShut {NoStop}%
\bibitem [{\citenamefont {Winkler}\ \emph {et~al.}(1989)\citenamefont {Winkler}, \citenamefont {Kotthaus},\ and\ \citenamefont {Ploog}}]{winkler_landau_1989}%
  \BibitemOpen
  \bibfield  {author} {\bibinfo {author} {\bibfnamefont {R.~W.}\ \bibnamefont {Winkler}}, \bibinfo {author} {\bibfnamefont {J.~P.}\ \bibnamefont {Kotthaus}},\ and\ \bibinfo {author} {\bibfnamefont {K.}~\bibnamefont {Ploog}},\ }\bibfield  {title} {\bibinfo {title} {Landau band conductivity in a two-dimensional electron system modulated by an artificial one-dimensional superlattice potential},\ }\href {https://doi.org/10.1103/PhysRevLett.62.1177} {\bibfield  {journal} {\bibinfo  {journal} {Physical Review Letters}\ }\textbf {\bibinfo {volume} {62}},\ \bibinfo {pages} {1177} (\bibinfo {year} {1989})},\ \bibinfo {note} {publisher: American Physical Society}\BibitemShut {NoStop}%
\bibitem [{\citenamefont {Mirlin}\ and\ \citenamefont {Wölfle}(1998)}]{mirlin_weiss_1998}%
  \BibitemOpen
  \bibfield  {author} {\bibinfo {author} {\bibfnamefont {A.~D.}\ \bibnamefont {Mirlin}}\ and\ \bibinfo {author} {\bibfnamefont {P.}~\bibnamefont {Wölfle}},\ }\bibfield  {title} {\bibinfo {title} {Weiss oscillations in the presence of small-angle impurity scattering},\ }\href {https://doi.org/10.1103/PhysRevB.58.12986} {\bibfield  {journal} {\bibinfo  {journal} {Physical Review B}\ }\textbf {\bibinfo {volume} {58}},\ \bibinfo {pages} {12986} (\bibinfo {year} {1998})},\ \bibinfo {note} {publisher: American Physical Society}\BibitemShut {NoStop}%
\bibitem [{\citenamefont {Kaganov}\ and\ \citenamefont {Slutskin}(1983)}]{kaganov_coherent_1983}%
  \BibitemOpen
  \bibfield  {author} {\bibinfo {author} {\bibfnamefont {M.~I.}\ \bibnamefont {Kaganov}}\ and\ \bibinfo {author} {\bibfnamefont {A.~A.}\ \bibnamefont {Slutskin}},\ }\bibfield  {title} {{\selectlanguage {english}\bibinfo {title} {Coherent magnetic breakdown}},\ }\href {https://doi.org/10.1016/0370-1573(83)90006-6} {\bibfield  {journal} {\bibinfo  {journal} {Physics Reports}\ }\textbf {\bibinfo {volume} {98}},\ \bibinfo {pages} {189} (\bibinfo {year} {1983})}\BibitemShut {NoStop}%
\bibitem [{\citenamefont {Eddy}\ and\ \citenamefont {Stark}(1982)}]{eddy_haas---van_1982}%
  \BibitemOpen
  \bibfield  {author} {\bibinfo {author} {\bibfnamefont {J.~W.}\ \bibnamefont {Eddy}}\ and\ \bibinfo {author} {\bibfnamefont {R.~W.}\ \bibnamefont {Stark}},\ }\bibfield  {title} {\bibinfo {title} {de {Haas}---van {Alphen} {Study} of {Coherent} {Magnetic} {Breakdown} in {Magnesium}},\ }\href {https://doi.org/10.1103/PhysRevLett.48.275} {\bibfield  {journal} {\bibinfo  {journal} {Physical Review Letters}\ }\textbf {\bibinfo {volume} {48}},\ \bibinfo {pages} {275} (\bibinfo {year} {1982})},\ \bibinfo {note} {publisher: American Physical Society}\BibitemShut {NoStop}%
\bibitem [{\citenamefont {Fortin}\ and\ \citenamefont {Ziman}(1998)}]{fortin_frequency_1998}%
  \BibitemOpen
  \bibfield  {author} {\bibinfo {author} {\bibfnamefont {J.-Y.}\ \bibnamefont {Fortin}}\ and\ \bibinfo {author} {\bibfnamefont {T.}~\bibnamefont {Ziman}},\ }\bibfield  {title} {\bibinfo {title} {Frequency {Mixing} of {Magnetic} {Oscillations}: {Beyond} {Falicov}-{Stachowiak} {Theory}},\ }\href {https://doi.org/10.1103/PhysRevLett.80.3117} {\bibfield  {journal} {\bibinfo  {journal} {Physical Review Letters}\ }\textbf {\bibinfo {volume} {80}},\ \bibinfo {pages} {3117} (\bibinfo {year} {1998})},\ \bibinfo {note} {publisher: American Physical Society}\BibitemShut {NoStop}%
\bibitem [{\citenamefont {Gvozdikov}\ \emph {et~al.}(2002)\citenamefont {Gvozdikov}, \citenamefont {Pershin}, \citenamefont {Steep}, \citenamefont {Jansen},\ and\ \citenamefont {Wyder}}]{gvozdikov_haas--van_2002}%
  \BibitemOpen
  \bibfield  {author} {\bibinfo {author} {\bibfnamefont {V.~M.}\ \bibnamefont {Gvozdikov}}, \bibinfo {author} {\bibfnamefont {Y.~V.}\ \bibnamefont {Pershin}}, \bibinfo {author} {\bibfnamefont {E.}~\bibnamefont {Steep}}, \bibinfo {author} {\bibfnamefont {A.~G.~M.}\ \bibnamefont {Jansen}},\ and\ \bibinfo {author} {\bibfnamefont {P.}~\bibnamefont {Wyder}},\ }\bibfield  {title} {\bibinfo {title} {de {Haas}--van {Alphen} oscillations in the quasi-two-dimensional organic conductor kappa {(ET)2Cu(NCS)2}: {The} magnetic breakdown approach},\ }\href {https://doi.org/10.1103/PhysRevB.65.165102} {\bibfield  {journal} {\bibinfo  {journal} {Physical Review B}\ }\textbf {\bibinfo {volume} {65}},\ \bibinfo {pages} {165102} (\bibinfo {year} {2002})},\ \bibinfo {note} {publisher: American Physical Society}\BibitemShut {NoStop}%
\bibitem [{\citenamefont {Al'tshuler}\ \emph {et~al.}(1981)\citenamefont {Al'tshuler}, \citenamefont {Aronov},\ and\ \citenamefont {Spivak}}]{altshuler_aaronov-bohm_1981}%
  \BibitemOpen
  \bibfield  {author} {\bibinfo {author} {\bibfnamefont {B.~L.}\ \bibnamefont {Al'tshuler}}, \bibinfo {author} {\bibfnamefont {A.~G.}\ \bibnamefont {Aronov}},\ and\ \bibinfo {author} {\bibfnamefont {B.~Z.}\ \bibnamefont {Spivak}},\ }\bibfield  {title} {\bibinfo {title} {The {Aaronov}-{Bohm} effect in disordered conductors},\ }\href@noop {} {\bibfield  {journal} {\bibinfo  {journal} {JETP Letters}\ }\textbf {\bibinfo {volume} {33}},\ \bibinfo {pages} {101} (\bibinfo {year} {1981})}\BibitemShut {NoStop}%
\bibitem [{\citenamefont {Pippard}(1962)}]{pippard_quantization_1962}%
  \BibitemOpen
  \bibfield  {author} {\bibinfo {author} {\bibfnamefont {A.~B.}\ \bibnamefont {Pippard}},\ }\bibfield  {title} {\bibinfo {title} {Quantization of coupled orbits in metals},\ }\href {https://doi.org/10.1098/rspa.1962.0200} {\bibfield  {journal} {\bibinfo  {journal} {Proceedings of the Royal Society of London. Series A. Mathematical and Physical Sciences}\ }\textbf {\bibinfo {volume} {270}},\ \bibinfo {pages} {1} (\bibinfo {year} {1962})},\ \bibinfo {note} {publisher: Royal Society}\BibitemShut {NoStop}%
\bibitem [{\citenamefont {Groth}\ \emph {et~al.}(2014)\citenamefont {Groth}, \citenamefont {Wimmer}, \citenamefont {Akhmerov},\ and\ \citenamefont {Waintal}}]{Groth2014}%
  \BibitemOpen
  \bibfield  {author} {\bibinfo {author} {\bibfnamefont {C.~W.}\ \bibnamefont {Groth}}, \bibinfo {author} {\bibfnamefont {M.}~\bibnamefont {Wimmer}}, \bibinfo {author} {\bibfnamefont {A.~R.}\ \bibnamefont {Akhmerov}},\ and\ \bibinfo {author} {\bibfnamefont {X.}~\bibnamefont {Waintal}},\ }\bibfield  {title} {\bibinfo {title} {{Kwant}: a software package for quantum transport},\ }\href {https://iopscience.iop.org/1367-2630/16/6/063065/article} {\bibfield  {journal} {\bibinfo  {journal} {New J. Phys.}\ }\textbf {\bibinfo {volume} {16}},\ \bibinfo {pages} {063065} (\bibinfo {year} {2014})}\BibitemShut {NoStop}%
\bibitem [{\citenamefont {B\"uttiker}(1986)}]{Buttiker1986}%
  \BibitemOpen
  \bibfield  {author} {\bibinfo {author} {\bibfnamefont {M.}~\bibnamefont {B\"uttiker}},\ }\bibfield  {title} {\bibinfo {title} {Four-terminal phase-coherent conductance},\ }\href@noop {} {\bibfield  {journal} {\bibinfo  {journal} {Phys. Rev. Lett.}\ }\textbf {\bibinfo {volume} {57}},\ \bibinfo {pages} {1761} (\bibinfo {year} {1986})}\BibitemShut {NoStop}%
\bibitem [{\citenamefont {Tkachenko}\ \emph {et~al.}(2022)\citenamefont {Tkachenko}, \citenamefont {Tkachenko}, \citenamefont {Baksheev},\ and\ \citenamefont {Sushkov}}]{Tkachenko2022}%
  \BibitemOpen
  \bibfield  {author} {\bibinfo {author} {\bibfnamefont {O.~A.}\ \bibnamefont {Tkachenko}}, \bibinfo {author} {\bibfnamefont {V.~A.}\ \bibnamefont {Tkachenko}}, \bibinfo {author} {\bibfnamefont {D.~G.}\ \bibnamefont {Baksheev}},\ and\ \bibinfo {author} {\bibfnamefont {O.~P.}\ \bibnamefont {Sushkov}},\ }\bibfield  {title} {\bibinfo {title} {Wannier diagrams for semiconductor artificial graphene},\ }\href {https://doi.org/DODOI:10.1134/S0021364022602020} {\bibfield  {journal} {\bibinfo  {journal} {JETP Letters}\ }\textbf {\bibinfo {volume} {116}},\ \bibinfo {pages} {638} (\bibinfo {year} {2022})}\BibitemShut {NoStop}%
\bibitem [{\citenamefont {Tkachenko}\ \emph {et~al.}(2023)\citenamefont {Tkachenko}, \citenamefont {Tkachenko}, \citenamefont {Baksheev},\ and\ \citenamefont {Sushkov}}]{Tkachenko2023}%
  \BibitemOpen
  \bibfield  {author} {\bibinfo {author} {\bibfnamefont {O.~A.}\ \bibnamefont {Tkachenko}}, \bibinfo {author} {\bibfnamefont {V.~A.}\ \bibnamefont {Tkachenko}}, \bibinfo {author} {\bibfnamefont {D.~G.}\ \bibnamefont {Baksheev}},\ and\ \bibinfo {author} {\bibfnamefont {O.~P.}\ \bibnamefont {Sushkov}},\ }\bibfield  {title} {\bibinfo {title} {Effect of disorder on magnetotransport in semiconductor artificial graphene},\ }\href {https://doi.org/DDOI:10.1134/S0021364022603219} {\bibfield  {journal} {\bibinfo  {journal} {JETP Letters}\ }\textbf {\bibinfo {volume} {117}},\ \bibinfo {pages} {222} (\bibinfo {year} {2023})}\BibitemShut {NoStop}%
\bibitem [{\citenamefont {Lifshits}\ and\ \citenamefont {Kosevich}(1956)}]{lifshits_theory_1956}%
  \BibitemOpen
  \bibfield  {author} {\bibinfo {author} {\bibfnamefont {I.~M.}\ \bibnamefont {Lifshits}}\ and\ \bibinfo {author} {\bibfnamefont {A.~M.}\ \bibnamefont {Kosevich}},\ }\bibfield  {title} {\bibinfo {title} {Theory of {Magnetic} {Susceptibility} in {Metals} at {Low} {Temperature}},\ }\href@noop {} {\bibfield  {journal} {\bibinfo  {journal} {Journal of Experimental and Theoretical Physics}\ }\textbf {\bibinfo {volume} {2}},\ \bibinfo {pages} {636} (\bibinfo {year} {1956})}\BibitemShut {NoStop}%
\bibitem [{\citenamefont {Landau}\ and\ \citenamefont {Lifshitz}(1980)}]{landau_chapter_1980}%
  \BibitemOpen
  \bibfield  {author} {\bibinfo {author} {\bibfnamefont {L.~D.}\ \bibnamefont {Landau}}\ and\ \bibinfo {author} {\bibfnamefont {E.~M.}\ \bibnamefont {Lifshitz}},\ }\bibfield  {title} {\bibinfo {title} {{CHAPTER} {VII} - {NON}-{IDEAL} {GASES}},\ }in\ \href {https://doi.org/10.1016/B978-0-08-057046-4.50014-2} {\emph {\bibinfo {booktitle} {Statistical {Physics} ({Third} {Edition})}}},\ \bibinfo {editor} {edited by\ \bibinfo {editor} {\bibfnamefont {L.~D.}\ \bibnamefont {Landau}}\ and\ \bibinfo {editor} {\bibfnamefont {E.~M.}\ \bibnamefont {Lifshitz}}}\ (\bibinfo  {publisher} {Butterworth-Heinemann},\ \bibinfo {address} {Oxford},\ \bibinfo {year} {1980})\ pp.\ \bibinfo {pages} {225--250}\BibitemShut {NoStop}%
\bibitem [{\citenamefont {Zener}\ and\ \citenamefont {Fowler}(1932)}]{zener_non-adiabatic_1932}%
  \BibitemOpen
  \bibfield  {author} {\bibinfo {author} {\bibfnamefont {C.}~\bibnamefont {Zener}}\ and\ \bibinfo {author} {\bibfnamefont {R.~H.}\ \bibnamefont {Fowler}},\ }\bibfield  {title} {\bibinfo {title} {Non-adiabatic crossing of energy levels},\ }\href {https://doi.org/10.1098/rspa.1932.0165} {\bibfield  {journal} {\bibinfo  {journal} {Proceedings of the Royal Society of London. Series A, Containing Papers of a Mathematical and Physical Character}\ }\textbf {\bibinfo {volume} {137}},\ \bibinfo {pages} {696} (\bibinfo {year} {1932})},\ \bibinfo {note} {publisher: Royal Society}\BibitemShut {NoStop}%
\bibitem [{\citenamefont {Whittaker}\ and\ \citenamefont {Watson}(1996)}]{whittaker_confluent_1996}%
  \BibitemOpen
  \bibfield  {author} {\bibinfo {author} {\bibfnamefont {E.~T.}\ \bibnamefont {Whittaker}}\ and\ \bibinfo {author} {\bibfnamefont {G.~N.}\ \bibnamefont {Watson}},\ }\bibfield  {title} {\bibinfo {title} {The {Confluent} {Hypergeometric} {Function}},\ }in\ \href {https://doi.org/10.1017/CBO9780511608759.017} {\emph {\bibinfo {booktitle} {A {Course} of {Modern} {Analysis}}}},\ \bibinfo {series and number} {Cambridge {Mathematical} {Library}}\ (\bibinfo  {publisher} {Cambridge University Press},\ \bibinfo {address} {Cambridge},\ \bibinfo {year} {1996})\ \bibinfo {edition} {4th}\ ed.,\ pp.\ \bibinfo {pages} {337--354}\BibitemShut {NoStop}%
\bibitem [{\citenamefont {Gradshteyn}\ and\ \citenamefont {Ryzhik}(2014)}]{zwillinger_9_2014}%
  \BibitemOpen
  \bibfield  {author} {\bibinfo {author} {\bibfnamefont {I.~S.}\ \bibnamefont {Gradshteyn}}\ and\ \bibinfo {author} {\bibfnamefont {I.~M.}\ \bibnamefont {Ryzhik}},\ }\bibfield  {title} {{\selectlanguage {english}\bibinfo {title} {9 - {Special} {Functions}}},\ }in\ \href {https://doi.org/10.1016/B978-0-12-384933-5.00009-6} {{\selectlanguage {english}\emph {\bibinfo {booktitle} {Table of {Integrals}, {Series}, and {Products} ({Eighth} {Edition})}}}},\ \bibinfo {editor} {edited by\ \bibinfo {editor} {\bibfnamefont {D.}~\bibnamefont {Zwillinger}}\ and\ \bibinfo {editor} {\bibfnamefont {V.}~\bibnamefont {Moll}}}\ (\bibinfo  {publisher} {Academic Press},\ \bibinfo {address} {Boston},\ \bibinfo {year} {2014})\ pp.\ \bibinfo {pages} {1014--1059}\BibitemShut {NoStop}%
\end{thebibliography}
\end{document}